\documentclass[article]{IEEEtran}

\usepackage{graphicx}
\usepackage{makecell}
\usepackage{multirow}
\usepackage[noadjust]{cite}
\usepackage[hyphens]{url}
\usepackage{pdflscape}
\usepackage{afterpage}
\usepackage{rotating}
\usepackage{color}
\usepackage{soul}
\usepackage[textsize=small,textwidth=1.2cm]{todonotes}

\begin{document}

\title{Will SDN be part of 5G?}

\author{\IEEEauthorblockN{Zainab Zaidi\IEEEauthorrefmark{1},
Vasilis Friderikos\IEEEauthorrefmark{2},
Zarrar Yousaf\IEEEauthorrefmark{3},
Simon Fletcher\IEEEauthorrefmark{4},
Mischa Dohler\IEEEauthorrefmark{2}
and Hamid Aghvami\IEEEauthorrefmark{2}\\}
\IEEEauthorblockA{\IEEEauthorrefmark{1}IEEE Senior Member, Email: zzaidi@ieee.org\\}
\IEEEauthorblockA{\IEEEauthorrefmark{2}Centre for Telecommunications Research\\
King's Collge London, Strand, London, WC2R 2LS, UK\\
Email: \{vasilis.friderikos, mischa.dohler, hamid.aghvami\}@kcl.ac.uk\\}
\IEEEauthorblockA{\IEEEauthorrefmark{3}NEC Laboratories Europe, Heidelberg, Baden-W$\ddot{\mbox{u}}$rttemberg, Germany\\
Email: Zarrar.Yousaf@neclab.eu\\}
\IEEEauthorblockA{\IEEEauthorrefmark{4} Real Wireless Ltd, Pulborough, RH20 4XB, UK\\
Email: simon.fletcher@realwireless.biz\\}
}

\maketitle
\renewcommand{\baselinestretch}{1.1}

\begin{abstract}
For many, this is no longer a valid question and the case is considered settled with SDN/NFV (Software Defined Networking/Network Function Virtualization) providing the inevitable innovation enablers towards the realization of a virtualized, flexible, programmable and flexible 5G network. As SDN along with other technology enablers (including NFV) are still in the process of evolution, the first commercial deployment of 5G may take few years. However, some companies are claiming the availability of 5G solutions, but considering the monumental task of softwarization of mobile cellular networks there are genuine concerns that we may only see some point solutions involving SDN technology instead of a fully virtualized SDN-enabled 5G mobile network. In order to determine the technology readiness of SDN solutions in the context of 5G networks, this survey paper attempts to identify all important obstacles in the way, and looks at the state of the art of the relevant research. This survey is different from the previous surveys on SDN-based mobile networks  as it focuses on the salient problems and discusses solutions proposed within and outside SDN literature. Our main focus is on fronthaul, backward compatibility, supposedly disruptive nature of SDN deployment, business cases and monetization of SDN related upgrades, latency of general purpose processors (GPP), and additional security vulnerabilities that softwarization brings along to the mobile network. We have also provided a summary of the architectural developments in SDN-based mobile network landscape, including deployment options for SDN within NFV framework, as not all work can be covered under the focused issues. This paper provides a comprehensive survey on the state of the art of SDN-based mobile networks and clearly points out the gaps in the technology.
\end{abstract}


\section{Introduction}
While 5G is still not standardized\footnote{The Non-standalone 5G NR (New Radio) specifications are recently approved in Dec.\ 2017 and the rest of Rel.\ 15 (Standalone 5G NR) has expected functional freeze date of Sep.\ 2018. The work on standardization of NFV has moved already to detailed specification and proof-of-concept by ETSI. IEEE is also developing standard for NGFI, a new fronthaul technology.}
, this paper is trying to point out that there is a significant chance that the idea of a fully programmable control plane with switches may not be tech-ready for 5G. {The emphasis of this paper is only on the mobile or cellular networks (both access and core networks), although, 5G may encompass several other technologies for connectivity, i.e., satellite, cable, optical, WLAN (Wireless LAN) etc.} 

The major focus of SDN technology is towards decoupling of the software-based control plane from the hardware-based data plane (e.g., packets forwarding) of networking and switching pieces of equipment. SDN brings with it unprecedented ease in innovation, openness, optimum resource utilization, support for virtualization, etc., but it is perhaps the most disruptive idea, the mobile networks have seen since their transition from 1G to 2G, i.e., from analog to digital systems. Even LTE did not come with the similar degree of disruption \cite{Fujitsu2014}. The disruption required may possibly push the SDN deployments, and as a consequence the development of SDN-based solutions, to further future. 5G may just see some point solutions with SDN in some places \cite{Fujitsu2014} and the focus would largely remain on densification, mm-wave technology, and DAS (Distributed Antenna Systems) to achieve 1000x capacity, 100x data rate, and 100x active connections from the already deployed LTE network. An interesting point here is that all of the salient targets and business opportunities defined for 5G are not strictly dependent on SDN. {Nevertheless, it is important to mention that 3GPP's 5G NR (New Radio) is a step closer to SDN-based mobile networks due to complete separation of control plane (CP) from user plane in NGC (New Generation Core), modularizing of CP functions, and defining the interfaces between CP functions as services {\cite{3gpp.23.2017}}.}
%
%
%
%
%
%

In recent years, there has been a lot of interest in SDN-based mobile network and there are a number of papers proposing SDN-based mobile networks architectures and listing the huge benefits they can bring to the mobile industry \cite{Gudipati.hotsdn13, Li.ewsdn12, Bhaumik.mobicom12, Bansal.hotsdn12, Yang.sigcomm13, Akyildiz2015, Lai2015, Raza.racs14, Liu2014}. These papers were basically advocating the case for SDN by emphasizing the positives, a very understandable trend to address the need of the time. As the time passes, researchers are also raising concerns about getting the technology ready for deployments and respective obstacles along the road. The state of the art of SDN technology in wired and fixed network is much more advanced than the SDN-based mobile networks developments. However, with 87\% of the total Internet users now in possession of a smart-phone\footnote{\url{http://www.globalwebindex.net/blog/87-of-internet-users-now-have-a-smartphone}}, mobility is very important for global telecommunication network. LTE Evolved Packet Core (EPC) has done a remarkable job in simplifying the core and separating control and user plane to an extent. {3GPP's 5G NGC has further improved this separation.} The base-station eNodeB, however, still contains both planes. Moreover, shared wireless medium and interference between neighboring base-station make it more difficult to realize separation of control from forwarding plane. 
%
%
%
%

We have identified six major obstacles and issues which need to be addressed in order for the technology to move forward. There are certainly other issues as well and as the technological development in this area progresses, there will be an even longer list of questions. We have selected the following six issues according to their relevance with research and innovation. Other issues, such as, re-organization of telecommunication industry to suit the SDN model, regulatory issues, etc., are outside the scope of the present paper. Nonetheless, they are equally important issues to be discussed by the SDN community. 
{Moreover, other issues related to optimization, added functionalities, etc., are also outside the scope of present paper. Examples of such issues are energy efficiency of SDN networks, optimized resources consumption, intelligent networking techniques for big data, integration of IoT (Internet of Things), etc. {\cite{Wu2016}}. However, these issues are also important, e.g., the additional energy and resource consumption introduced by SDN and NFV should also be considered and efficient techniques should be developed for greening the future networks {\cite{Wu2016}}. The major obstacles are:}

\begin{enumerate}
\item fronthaul;
\item latency of general purpose platforms;
\item backward compatibility;
\item disruptive deployment;
\item SDN specific security vulnerabilities;
\item clear and compelling business case.
\end{enumerate}

We explain the issues in detail in Section \ref{issues} and they set the standard according to which we survey the existing proposals of SDN-based mobile networks. It is very interesting to see that although there are many publications regarding the new architecture, very few touch upon the real hurdles. Although, there has been significant efforts to explore solutions for the above mentioned issues but they are sporadic and isolated and a comprehensive solution is still missing. 
{The main motivation behind this survey paper is to understand the state of the art and highlight the gap between what is being done versus what is needed to be done. A big picture or broad perspective is a compass to help us steer towards the goal. Although, it is not trivial to have a big picture for such a complex problem showing all significant dimensions of the problem space. Major obstacles and issues in the path of technology deployment is one way of constructing the big picture which is meaningful and very effective in understanding the state of the art.} 
As per our knowledge, none of the surveys on SDN-based mobile networks has looked into the matter of obstacles as the main focus. In Section \ref{rel_work}, we have presented a brief account of available surveys on SDN-based mobile networks proposals and how our paper is different from them.

Once again, we want to emphasize that this paper is not disputing the benefits of SDN rather we consider them substantial and necessary for the future generations of networks. Our question is not "why?", it is "when?". Considering the gigantic task, it is vital to prioritize the research issues in order of their impact over the SDN technology's development and realization. 
{The main contribution of this survey paper is to provide a clearer picture to the readers about what is needed to be done regarding the realization of SDN-based mobile networks. We achieved the objective by:}
\begin{enumerate}
\item {identifying the major issues and obstacles which should be resolved so the technology could reach the deployment stage;}
\item {providing the state of the respective solutions, lessons learned, comparison of alternate proposals wherever possible, and highlighting the gaps through summaries, tables and lists;}
\item {pointing out the standardization activities wherever possible;}
\item {surveying the architectural development in SDN-based mobile networks around the popular themes to get an understanding of the outlook of future networks.}
\end{enumerate}

The paper is organized as follows: Section \ref{background} contains a brief background about SDN, NFV, CloudRAN, and current and future LTE architectures. This section is added to make this paper self-contained, readers knowledgeable in the above technologies can easily skip this section and go directly to Section \ref{issues}. As discussed above, Section \ref{issues} discusses the major obstacles and issues in detail and also lists the solutions proposed in the literature. Section \ref{sdran} then summarizes the SDN-based mobile networks literature as the current state of the art depicting specially the achievements unrelated to the obstacles and issues identified in this paper. Section \ref{rel_work} discusses the SDN-based mobile networks' surveys and what was the need to have the present one. Finally Section \ref{conc} summarizes the major take-away points made in this paper.

\section{Background}\label{background}

In this section, we present brief discussions about SDN, NFV, CloudRAN and LTE including some future proposals for mobile network architecture.    

\subsection{SDN}

\begin{figure}
\begin{center}
\includegraphics[scale = 0.4, clip=true]{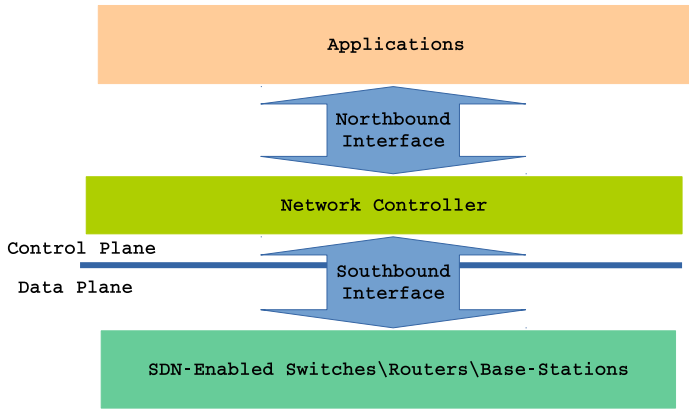}
\caption{A generic SDN model of three layers.}
\label{SDN-model}
\end{center}
\end{figure}

SDN mainly focuses towards decoupling of the software-based control plane from the hardware-based data plane (e.g., packets forwarding) of networking and switching pieces of equipment. The basic SDN model is shown in Fig.\ \ref{SDN-model}. The logically centralized controllers contain the control logic to translate the application requirements down to the data plane and are responsible in providing an abstract network view to the application plane. The major issue is to create appropriate mapping of the existing network functionalities to the decoupled control and forwarding planes. Conforming to the terminologies used in the research community, the interface south of control plane is called the southbound interface in Fig.\ \ref{SDN-model} and north of control plane is the northbound interface. An important point to note here is that both interfaces define the bi-directional information flow contrary to what may be implied by their names. 
%
%

\begin{figure}
\begin{center}
\includegraphics[scale = 0.45, clip=true]{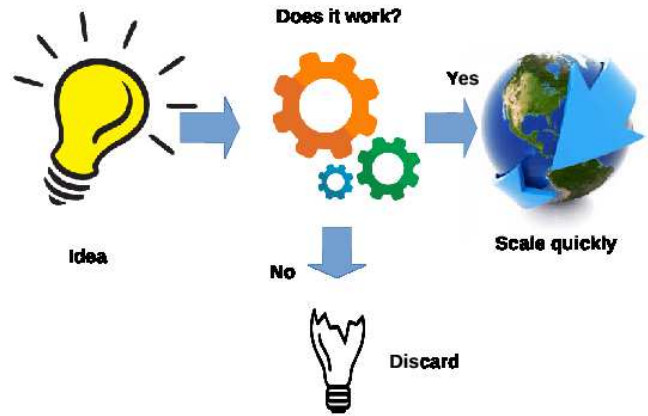}
\caption{The fail-fast model, major benefit of SDN technology.}
\label{fail-fast}
\end{center}
\end{figure}

The major benefit of softwarization is flexible innovation as other advantages come indirectly, such as, efficient resource utilization is a result of centralization not softwarization. Also, novel business opportunities through RAN-as-a-service (RaaS) is a possibility through cloud computing and virtualization which not necessarily need softwarization. Similarly, NFV is a complementary technology to SDN and does not depend on it \cite{Fujitsu2014}. Flexible innovation or fail-fast approach, as shown in Fig.\ \ref{fail-fast}, is the important feature which is not possible without SDN. In fail-fast model, ideas are tried through rapid deployment, discarded if failed, and scaled out quickly if taken off. Currently, any upgrade in mobile network takes a long time. With SDN, any mobile technology will have much smaller trial and deployment process. 

The LTE architecture also distinguishes between user-plane and control-plane, where the former deals with data packet forwarding, while the latter focuses on signaling and management messages and operations using the same physical infrastructure. Both planes reside in the firmware of the system. This demarcation is much clearly designed into the mobile core network EPC (Evolved Packet Core), but the radio access part of LTE consists of only base-station node, eNodeB, performing data forwarding and control functions. The SDN-based core network requires transporting the control plane into software along with the control logic required for the data forwarding plane, e.g., routing rules, mobility anchoring, etc. We also remark that user plane as defined by LTE is not exactly the same as data forwarding plane of SDN and similarly both control planes also differ slightly \cite{Zaidi2015}. 
%
%

\subsection{NFV}

\begin{figure}
\begin{center}
\includegraphics[scale = 0.39, clip=true]{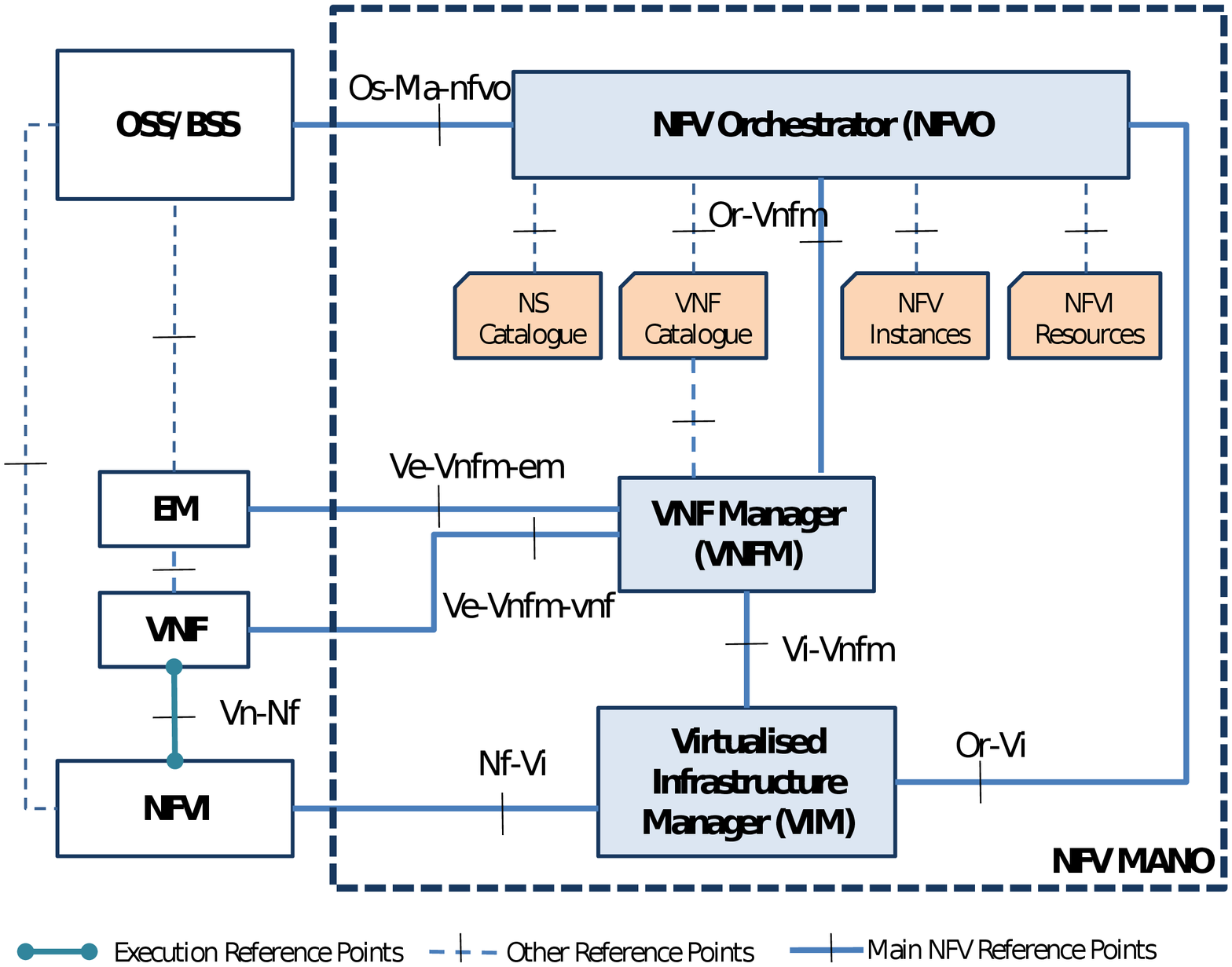}
\caption{ETSI NFV Management and Orchestration (MANO) Framework Overview.}
\label{fig:mano}
\end{center}
\end{figure}

In NFV (Network Function Virtualization), network functions are implemented in software components called virtual network functions (VNFs) \cite{Abdelwahab2016}. VNFs are deployed on commodity servers or cloud infrastructure instead of dedicated hardware reducing CAPEX (CAPtial EXpenditure) and optimizing resource provisioning \cite{Abdelwahab2016}. The VNFs might not only be linked to a specific service and/or application but they can relate to specific user requirements, network policies and allow for service differentiation based on the user subscription SLAs (Service Level Agreements). The key motivation is that once implemented in software these VNFs are able to run in general-purpose hardware where they can be deployed, configured, scaled, migrated, and updated on demand. Operation of the VNFs is expected to take place in accordance to the ETSI Network Function Virtualization (NFV) management and organization (MANO) framework depicted in Fig.\ \ref{fig:mano} \cite{etsi-mano}.

%

%
%

The MANO framework defines three main logical components, namely the Virtualized Infrastructure Manager (VIM), the VNF Manager (VNFM) and the NFV Orchestrator (NFVO), interconnected over specific reference points. The VIM manages the infrastructure resources (e.g., compute, network, storage, memory) whereas the VNFM performs the life-cycle management of the individual VNFs (such as scaling, migrating, update/upgrade etc). The NFVO is responsible for both the resource/service orchestration of a Network Service (NS), which is composed of one or more VNFs (and its components) interconnected over virtual links. There are additional data repositories that may contain necessary information about NS, VNF, NFV and NFVI (NFV Infrastructure) that will enable the NFVO to perform its tasks. The MANO architecture also defines reference points for interfacing the MANO system with external entities like NFVI, OSS/BSS, VNFs and Element Managers (EM) for delivering a unified management and orchestration of a VNF system.


%
%
%
%

SDN is a complementary technology to NFV and the realization of NFV does not depend on SDN. 
%

\subsection{LTE Architectures- current and future}\label{lte-background}

\begin{figure}
\begin{center}
\includegraphics[scale = 0.35, clip=true]{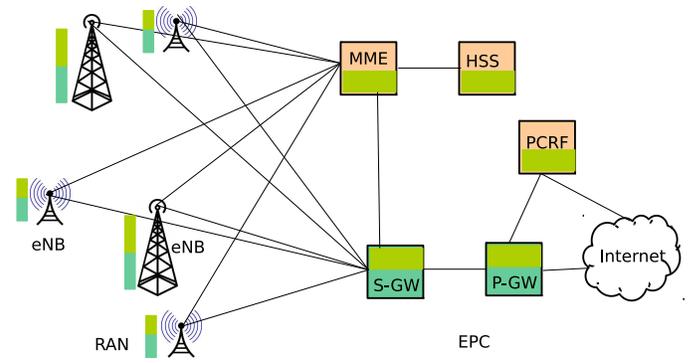}
\caption{LTE, RAN and core (EPC) network. The colors indicate the SDN functional decomposition: orange color shows application plane, green refers to control plane functions, and turquoise shows data plane. Solid lines show wired connections.}
\label{lte}
\end{center}
\end{figure}

The LTE architecture \cite{Alcatel.09}, shown in Fig.\ \ref{lte}, also distinguishes user plane, dealing with the data packet forwarding, and control plane, focusing on signaling and management messages and operations, using the same physical infrastructure. EPC (Evolved Packet Core) has the designated control plane nodes MME (Mobility Management Entity), HSS (Home Subscriber Server), and PCRF (Policy Control and Charging Rules Function) \cite{Alcatel.09}. Serving gateway (S-GW) and packet gateway (P-GW) are used for the data forwarding. From SDN's perspective, the gateways and the base-station also contain control plane functionality along with the designated control nodes of EPC. This is shown by color coding all EPC nodes and eNBs (eNodeB, base-station) partially with green color in Fig.\ \ref{lte} relating them to the control plane in Fig.\ \ref{SDN-model}. The base-stations eNBs and S/P-GWs are comprised of the data plane from SDN's point of view and thus shown with partial turquoise coloring in Fig.\ \ref{lte}. A possible implementation of EPC control nodes in a SDN environment is as application modules so we color them partially orange. The colors are relating the nodes in Fig.\ \ref{lte} to the layers in SDN model of Fig.\ \ref{SDN-model}. As it is clear from Fig.\ \ref{lte}, the traditional networks have control and forwarding planes integrated into all elements and a proper logical decomposition, while preserving the performance constraints over the inter-communication between controller and data plane elements, is the biggest challenge.      

In this section, we also discuss some of the ideas proposed for future mobile network architecture, i.e., Beyond Cellular Green Generation (BCG2) \cite{Rittenhouse.wcnc12} and Phantom cell concept \cite{Ishii.globecom12}. These ideas have shown big promises in energy efficiency \cite{Earthd3.3}. The logical decoupling of data transmissions and control signaling paradigm is one of the key directions being explored by GreenTouch\footnote{\url{www.greentouch.org}} under the project Beyond Cellular Green Generation (BCG2) \cite{Rittenhouse.wcnc12}.  In BCG2 architecture, the signaling nodes are responsible for the coverage and are usually assumed to deliver low rate services, such as, random access and paging, over long ranges; whereas the data nodes can be activated and deactivated depending on the traffic demand and it is designed for high rate and small ranges. The decoupling is logical in nature and a single location can host both types of nodes. A set of studies regarding the BCG2 architecture is performed under the EU FP7 IP project EARTH \cite{Earthd3.3}. The study shows that up to 85-90\% saving potential is possible with this revolutionary changed architecture compared to the current systems. Coverage is also separated from data processing in a CloudRAN architecture  \cite{Bhaumik.mobicom12,Aricent2016}, where a centralized BBU pool serves several RRH (Radio Remote Heads) in the area and not-in-service BBUs can be put in sleep mode to save energy. 

Phantom cell concept \cite{Ishii.globecom12} which was introduced for realizing true potential of dense deployment of small cells as suggested for LTE Release 12. In this idea, many small cells, called the phantom cells as they contain only LTE user plane, are overlaid with a normal macro cell which provides interference coordination. However, a macro cell consumes over 100 times more power than a pico/femto cell as described in the EU FP7 EARTH project \cite{Earthd3.3} and keeping it on all the time would lead to severe power inefficiency. \cite{Zaidi2015} proposed to change the macro base-station in Phantom cell architecture with a signaling only large coverage base-station which can also hold some control plane processing for SDN-based mobile network. The architecture can provide interference coordination service to the small base-stations in the area.

Although not yet explored, but the phantom cell or the data cell in BCG2 architecture can become a starting point to realize dump base-stations in a SDN-based mobile network architecture. The dump and small base-stations can provide data forwarding only and their control lies elsewhere, e.g., in the cloud or in the macro base-station. Striping down signaling capability from dump base-stations opens up bigger challenges as how to determine the best small base-station to connect to UE. Moreover, a full SDN-based mobile network can also allow us to choose the base-stations to remain on for signaling instead of a predetermined signaling base-station in BCG2. In a CloudRAN type architecture, the controller can select a set of RRHs to provide signaling functions and the rest RRHs can be used on-demand.

\subsection{CloudRAN}

\begin{figure}
\begin{center}
\includegraphics[scale = 0.35, clip=true]{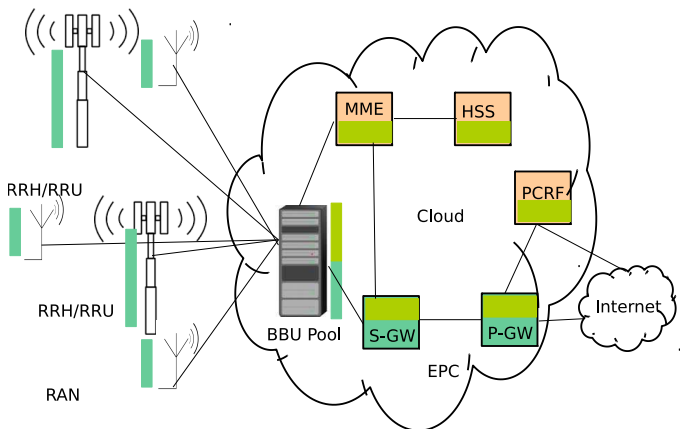}
\caption{The CloudRAN architecture, with BBU pool and EPC in the cloud. Colors indicate the SDN functional decomposition: orange shows application plane, green refers to control plane, and turquoise shows data plane. Solid lines show wired connections.}
\label{cran}
\end{center}
\end{figure}

CloudRAN has its roots firmly planted by China Mobile\footnote{\url{http://labs.chinamobile.com/cran}} and it can be viewed as a facilitator for introducing SDN technology. CloudRAN, which is now hitting its stride, centralizes all functionalities, control as well as data plane, into a centralized BBU (Base Band Unit) pool, or a data center/cloud, for easier management and coordination as shown in Fig.\ \ref{cran}.  In that respect, only antennas and some active RF components, i.e., RRU (Remote Radio Unit) or RRH (Remote Radio Heads), are left on the cell sites \cite{Bhaumik.mobicom12,Aricent2016}. Figure \ref{lte} is morphed into CloudRAN architecture by replacing eNB with RRHs/RRUs and moving the processing units, i.e., BBUs and the EPC to the cloud as shown in Fig.\ \ref{cran}. CloudRAN also does not decompose control and forwarding layers as shown by the color coding in Fig.\ \ref{cran} to match with SDN layers of Fig.\ \ref{SDN-model}. Although, the RRHs/RRUs can be classified as the data forwarding RF devices without any control functionality.

CloudRAN is proposed as a mechanism to realize small cell deployment in LTE through proper coordination for interference management. LTE small cell, however, assumes distributed control with self-organizing (SON) capabilities. The connection from the cell site to the serving gateway (S-GW) in LTE could be through wired or wireless links (http://scf.io/). On the other hand, CloudRAN assumes a high bandwidth fiber link between RRU and the centralized data center, which is also the most prohibitively expensive aspect of this proposal. 

Centralized processing in the cloud is not necessary for SDN and alternative proposals have been proposed to reduce the load on the fronthaul by leaving some processing blocks at the base-station \cite{Akyildiz2015}. Also, some proposals have left the part of control plane at the base-station in order to perform delay-critical control \cite{Gudipati.hotsdn13}. These proposals require connections between control elements in the cloud and at the base-stations.

{The CloudRAN architecture has the advantage of lower cost, energy efficiency, smaller footprint, and lower maintenance cost {\cite{Wu2012}}. Alcatel Lucent's Light Radio, Ericsson's Antenna Integrated Radio and Nokia Siemens Networks' Liquid Radio are among the earlier CloudRAN products.}

{The CloudRAN architecture, with its low cost and flexibility of operation, also opens up new possibilities. The architecture is shown to be extremely useful for very high-speed train communications\footnote{\url{https://www.globalrailnews.com/2017/12/04/samsung-successfully-tests-5g-on-high-speed-train-in-japan/}} {\cite{Luo2012}}. Due to low cost, several remote radio heads (RRHs) can be deployed along the track in the framework presented in {\cite{Luo2012}}. Centralized processing allows to use virtualization along with joint beamforming techniques to yield up to 100Mbps at the speed of 450kmph {\cite{Luo2012}}.} 

\section{Major Obstacles and Issues}\label{issues}

We identify six major issues hindering the realization of SDN technology. In no particular order, these issues are
\begin{enumerate}
\item Fronthaul
\item Latency of general purpose platforms
\item Backward compatibility
\item Disruptive deployment
\item SDN specific security vulnerabilities
\item Clear and compelling business case
\end{enumerate}
In this section these issues are discussed in detail. We have also looked into the state of the solutions proposed for them. This is one of the most rapidly changing field in digital technology and last few years have seen massive progress in the area. We cannot claim to have looked at each and every publication under the relevant themes but we have certainly covered the major solutions and ideas presented in the literature including industry white papers and technical reports as well as academic research papers.

\subsection{Fronthaul}

In centralized RAN architectures, such as, CloudRAN, connections from remote radio units (RRU) to the centralized baseband processing units (BBU) form fronthaul as opposed to backhaul, which connects the base-stations to the core network. RRUs are the wireless transceivers mainly consisting of the antenna heads. 

Traditionally, data from the processing cabinets of base-stations, in the form of digitized baseband signal, is carried to the antenna heads located at the rooftop or top of the masts by CPRI (Common Public Radio Interface) protocol over fiber optic connections. CPRI over fiber  allowed the antennas to be located away from the base-station cabinets as the legacy co-axial cables caused a lot of power loss and were also expensive and bulky \cite{Murphy2015} and can only be used for short distances. CPRI over fiber can connect RRUs miles away from BBUs with minimal losses, however, the limiting factor is the latency constraint to maintain channel state information and synchronization between BBU and RRU \cite{Murphy2015}.

The data to and from the centralized BBUs is digital version of high frequency baseband signal. As an example, a LTE baseband signal with 20MHz channel bandwidth needs to be sampled 30.72 million times per second for proper digitization resulting in data rate of approximately 2.5Gbps if data stream is duplicated for $2\times2$ MIMO \cite{Murphy2015}.  CPRI rate ranges from 614Mbps to 12Gbps \cite{Murphy2015}. Such high data rates need fiber optic cables, although, microwave can be used but for lower rate CPRI only. Interestingly, CPRI is very inefficient from data transport point of view, e.g., a 20MHz LTE channel can carry up to 150Mbps in downlink but requires CPRI rate of 2.5Gbps \cite{Murphy2015}. CPRI was not designed to carry data over very long distances and it was an internal base-station interface to connect the antennas on the rooftop to the base-station cabinet. There is clearly room for introducing efficiency in data transport in fronthaul.  Advanced techniques exploiting correlation among RRUs, such as, data compression, quantization, precoding, etc., can be used as well to reduce the data rate for the constraint fronthaul \cite{Peng2015}.

Other important issues for fronthaul are latency and jitter. Real-time communication, such as, VoLTE (voice over LTE), IoT, tactile applications, etc., and the control signaling between RRH and BBU to maintain channel state information require low delays. To keep one way delay under 75 microseconds, the length of the fiber link between RRH and BBU should not be more than 15Km \cite{Murphy2015}. According to \cite{Chitimalla2017}, the delay of CPRI should remain under 100${\mu}s$ and jitter should be under 65$ns$. Jitter in the synchronization signal between BBU and RRU will induce phase noise which in turn will lead to degradation in the transmitted modulated signal \cite{Gomes2015}. Heterogeneous CloudRAN \cite{Peng2014} tried to solve the latency issue by introducing a different high power node (HPN) than RRUs which is responsible for control signaling in a geographical area. However, the latency issue with data transmissions, such as, VoLTE, IoT, and other time-critical applications remains.

In fog/edge computing architectures \cite{OpenFog2016}, the nodes are widely spread, away from the cloud or data center, where the users are. They can provide some localized services and can also store some information. Such architecture also reduces the load on fronthaul with some improvement in latency requirements.

An effort to move away from semi-proprietary CPRI has also been made by ETSI through their Open Radio Equipment Interface (ORI) specifications\footnote{\url{http://www.etsi.org/technologies-clusters/technologies/ori}}. The ORI interface was built on top of CPRI and it supports line bit rates up to 10.14Gbps. The objective was to make the interface fully inter-operable and improving the data transport efficiency was not the main goal.

Among the various SDN-based mobile network architectures proposed in the literature, some have tried to touch upon fronthaul but not really looking into resolving the transport efficiency or latency issues. As shown in this section, the fronthaul is a big research problem in itself and we should not expect that research efforts towards new architectural organizations for SDN-based mobile networks would also be able to resolve fronthaul issues as well. But since fronthaul is a major element of SDN-based mobile network architectures, it cannot be left behind. Interestingly, we found only two papers \cite{Akyildiz2015,Gudipati.hotsdn13} which addressed the issue in their architectural design. Details about the design is discussed later in Section \ref{split-fronthaul}. The only other papers which discussed fronthaul in their architecture is CONCERT \cite{Liu2014} and SDCN \cite{Lai2015} but the major relevant issues were not their focus. CONCERT and SDCN proposed that fronthaul network can also be a SDN network with optical switches controlled and virtualized by SDN controller. CONCERT claimed virtualization efficiency with SDN fronthaul than using packet switched Ethernet \cite{Liu2014}.

\subsubsection{Functional split between BBU and RRU}\label{split-fronthaul}

\begin{figure}
\begin{center}
\includegraphics[scale = 0.35, clip=true]{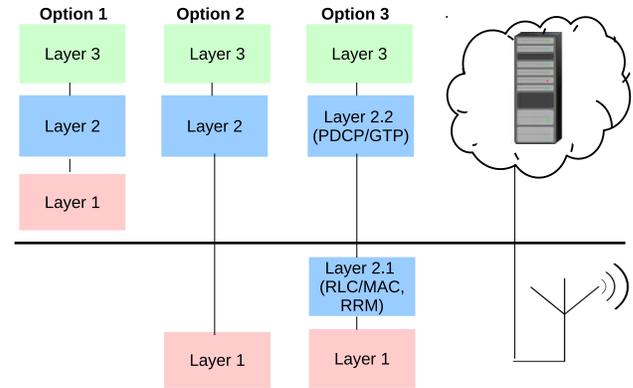}
\caption{Fronthaul functional split options (courtesy \cite{Aricent2016}).}
\label{f-split}
\end{center}
\end{figure}

Different centralization options are also considered to reduce the data rate for fronthaul \cite{Peng2015, Aricent2016} as shown in Fig.\ \ref{f-split}. When processing for all layers is performed in the centralized BBUs, it is called full centralization as shown by option 1 in Fig.\ \ref{f-split}. In this case, we have very high data rate waveform to transmit to and from the RRUs. Partial centralization options are also available where layer 1 or PHY was kept at the RRUs as it reduces a lot of fronthaul overhead \cite{Peng2015}. This is shown as option 2 in Fig.\ \ref{f-split}. 

Moreover, it sounds desirable to have all programmable OSI layers but there is a strong consensus among network owners and operators not to allow low level network programming by third parties \cite{Fujitsu2014} due to network security and management issues. Also, keeping layer 1 on dedicated hardware instead of general purpose processors can also improve the latency issue which is very critical for some IoT, tactical and real time applications. On the other hand, efficient implementation of some advanced features, such as, CoMP (Coordinated Multiple Point) and cooperative processing for massive MIMO will not be possible if layer 1 remains in the hardware. Interaction between layer 1 and 2 will also be complex and one way latency between RRU and BBU should be of the order of 100-150 microseconds \cite{Aricent2016}. Moreover, the fronthaul should also cater for some control signaling regarding carrier frequency, transmitter power, etc.

Another option for partial centralization is to split MAC function between the cloud and the antenna site. The one-way latency constraint can be more than a millisecond. This alternative is shown by option 3 in Fig.\ \ref{f-split}. In this architecture, HARQ of MAC is kept at RRU whereas the MAC scheduler resides in the central location \cite{Aricent2016}. In another option, only layer 3 resides in the cloud and PHY and MAC are kept at the antenna site \cite{Aricent2016}. This also reduces a lot of fronthaul overhead and handover overhead is also greatly reduced in case of mobility within the serving area of layer 3 as all cells can be aggregated and provided a single S1 view towards core network \cite{Aricent2016}.

Chinamobile presented NGFI (Next Generation Fronthaul Interface) \cite{Chih-Lin2015} which decouples the dependency of CPRI on number of antenna elements by putting all antenna related functionality, such as, downlink antenna mapping, fast Fourier transform [FFT], channel estimation, and equalization in RRU. It is shown that an LTE fronthaul bandwidth may decrease on the order of 100Mbps no matter how many antennas are used \cite{Chih-Lin2015}. 
{ Currently, NGFI is being standardized under IEEE 1914.1 project.} 

Similar ideas of doing cell based processing, such as, FFT/IFFT, subcarrier mapping, etc., of the receiving waveform in RRUs were discussed in \cite{Nikaein.mcs15}. According to the authors, this split will half the fronthaul capacity requirement and further reduction would be possible by moving demodulation and decoding blocks to the RRU. They also suggested to put full PHY for transmitting signal in the RRU to reduce fronthaul capacity. 

{Similar ideas of performing FFT/IFFT near the users is discussed and evaluated in {\cite{Thyagaturu2017}}, where LTE and DOCSIS (Data Over Cable Service Interface Specification) share the fibre fronthaul and remote FFT node. The remote FFT/IFFT reduces the fronthaul bit rate to approximately 1/30$^{\mbox{th}}$ of the conventional baseband approach. The system also used frequency domain I/Q symbols instead of conventional time domain I/Q symbols to reduce the required transport bit rate and used caching of the repetitive QAM symbols to further reduce the downstream bit rate requirement. Evaluations show 6.5\% (for full load) to 41\% (for 10\% load) savings for LTE traffic.}

SoftAir \cite{Akyildiz2015} is one of the two research papers on SDN-based mobile network architecture which also addressed fronthaul transport efficiency issue. In SoftAir modem (modulation and demodulation) is recommended to be kept at RRU in order to reduce the data rate of CPRI supported fronthaul. Such a split is also evaluated through hardware testbed experimentations in \cite{Mountaser.wcnc17} and latency and jitter are found to be within acceptable bound. The fronthaul is using Ethernet technology, although, the length of the fronthaul is not reported in \cite{Mountaser.wcnc17}. 

SoftRAN \cite{Gudipati.hotsdn13} provides another type of functional split, in terms of control functions, between center and remote site. They also realized that the biggest challenge in realizing SDN-based mobile network is the inherent delay between any centralized controller and the radio elements. The delay could typically be 5-10ms. SoftRAN splits the control plane functionalities and leaves the time-critical local control related to channel state, such as, downlink resource allocation (with transmit power set by the centralized controller), at the individual radio element.

A detailed evaluation of different functional splits is provided in \cite{Checko2016} in terms of multiplexing gains. The obvious conclusion is the best multiplexing gain for full centralization solution with all processing in the BBU pool. This is specially true for high load. When the traffic is low, the fronthaul requirements are shown to be relaxed with only higher layers in the BBU pool and lower layer processing near the RRU \cite{Checko2016}. The impact of functional split of fronthaul on its performance is studied with packetization and scheduling in \cite{Chang.icc16} and \cite{Chang.globecom16} and it is shown that there is a strong interplay between packetization overhead and latency that changes the fronthaul performance. For each split, the analysis finds the maximum number of supported RRUs for the best packetization method satisfying fronthaul capacity and latency constraints \cite{Chang.icc16}. This work exploits packetization latency and options of putting more processing at RRU to support more RRUs in the system for specific fronthaul constraints. The follow-up paper \cite{Chang.globecom16} also includes different scheduling mechanisms in the study to improve the number of supported RRUs.

A summary of split functionality proposals are presented in the Table \ref{Split_functions}, where it is clear that transport efficiency can be improved with proper distributions of functions between BBUs and RRUs. Although, no study has been done to compare the different proposals and we have used the generic terms 'Heavy' and 'Light' to show that all the split function proposals have lesser load than CloudRAN. It is not known how they compete with each other. In Table \ref{Split_functions}, RH stands for radio heads or transceivers and RRC is Radio Resource Control. Also, SoftAir \cite{Akyildiz2015} and NGFI \cite{Chih-Lin2015} did not mention any delay figure explicitly and we have taken $\leq$75$\mu$s as it is considered as a standard for centralized RAN \cite{Murphy2015}. The split architectures of SoftAir and NGFI focus on only reducing the load of fronthaul. In SoftAir, delay is considered as a constraint while forming clusters of RRUs. 

{\renewcommand{\arraystretch}{1.2}
\begin{center}
\begin{table*}
\centering
\small
\caption{Summary of split functions.}\label{Split_functions}
\begin{tabular}{ l| l l c c}
\hline
 Proposals & Cloud & RRU & Fronthaul load & Latency\\
\hline 
CloudRAN \cite{Aricent2016} & All layers & RH & Heavy &$\leq$75$\mu$s  \\  
NGFI \cite{Chih-Lin2015} & Rest  & Antenna related & Light & $\leq$75$\mu$\\
Partial/layer1 \cite{Aricent2016} & Layer 2+ & PHY/RH & Light& 100-150$\mu$s\\
Split MAC \cite{Aricent2016}  & RLC/Layer3+ & HARQ/PHY/RH & Light &~1ms\\
SoftAir \cite{Akyildiz2015} & PHY(ex. modulation)/Layer2+&Mod./Demod./RH & Light & $\leq$75$\mu$s \\
SoftRAN \cite{Gudipati.hotsdn13} & All layers (ex. RRC) &RRC/RH&Heavy&5-10ms\\
\hline
\end{tabular}
\end{table*}
\end{center}
}

\subsubsection{Ethernet vs.\ CPRI/ORI}

The use of Ethernet for fronthaul is also been looked at with various options of placing CPRI data or RF data directly into Ethernet frames \cite{Gomes2015}. Ethernet is a mature technology with standardized OAM (Operations, Administration, and Management) capabilities but by putting the data in Ethernet frame, the synchronization information could be lost \cite{Gomes2015}. Synchronous Ethernet, where clock information is extracted from the received data, and Precision Time Protocol, where time-stamped packets are exchanged, can be used to achieve synchronization in fronthaul Ethernet operation. 
{ Moreover, the traditional Ethernet switching cannot fulfill the strict timing requirements of fronthaul traffic and substantial optimization to the switching fabric is required {\cite{Oliva2015}}.

IEEE 802.1CM Time-Sensitive Network task group is looking into the set of standards focusing on time-synchronized low latency high bandwidth services over Ethernet networks for fronthaul transport network. In a centralized RAN, strict requirements are defined between the RRU and the BBU, such as, less than 0.1ms one-way latency at 614.4 to 24330.24Mbps {\cite{Lanzani2016}}. The standardization activity looks into techniques to reduce latency and jitter, such as prioritizing the time-sensitive frames, time-synchronization, end-to-end resource reservation, etc.}

In LTE small cell network, a massive coverage of backhaul is needed instead of fronthaul in centralized RAN and it is still considered one of the major issues 
{{\cite{Shakir2015}}. Both wired and wireless backhauls are being used according to the operating conditions. However, with excessive cost of lying fibre for dense networks and technological and regulatory issues with millimeter-wave backhauls, it is still an open problem in small cell network similar to fronthaul in CloudRAN type access networks. Smart backhaul/fronthaul solutions where interworking is exploited and joint optimization is performed with the access network are also recently explored {\cite{Shakir2015}}.  

A futuristic idea of reconfigurable backhaul/fronthaul, namely {\it Xhaul}, is presented in {\cite{Oliva2015}}. Here functional split between remote radio site and centralized processors depends on the network conditions and operational scenarios. A software defined xhaul employs different protocols, such as, CPRI/ORI, Ethernet over heterogeneous physical links, e.g., fibre, microwave, etc., with appropriate resource allocation to address latency, jitter, and throughput constraints for each type of data. 

Another proposal for software defined backhaul is presented in {\cite{Thyagaturu2016}}. Here a SDN orchestrator is responsible to manage multiple smart-gateways, a new element suggested in the scheme, in order to provide demand-based uplink capacity to the connected small cells. The smart-gateways are connected to the S/P-GW of multiple operators and allow resource sharing through SDN orchestrator.}

A significant advantage, in the case of backhaul, is the use of Ethernet instead of CPRI which has much better data transport efficiency. Moreover, commodity, or low-cost industry standard equipment can be used lowering the overall CAPEX where fiber is unavailable. 

{The most important development in this regard is IEEE 1914 working group (NGFI). The working group has two active projects: IEEE 1914.1 is studying NGFI of Chinamobile and IEEE 1914.3 is looking at encapsulation of digitized radio signal (I/Q samples) into Ethernet frames for fronthaul transmission. Also called as 'Radio over Ethernet', IEEE 1914.3 consolidates the efforts of a previous working group IEEE 1904. The Ethernet frame structure and MAC remains unchanged under this project and the focus is on encapsulation and mapping of digitized payload as well as control and management messages. 

An mmWave based fronthaul using Ethernet is presented in {\cite{Artuso2015}} for heterogeneous CloudRANs. The system contains small range RRUs and macro RRUs. An Ethernet switch is added with macro RRU to direct the traffic to the UE or to a connected small RRU. In this case, the small RRU does not need to decapsulate the Ethernet frame as it is done at macro RRU. Preliminary results show that the latency is within the limits for fronthaul operation.} 

In a recent study, an Ethernet based fronthaul is tested using hardware implementation  \cite{Mountaser.wcnc17} and latency and jitter are found to be within acceptable bound. The length of the fronthaul is not reported in \cite{Mountaser.wcnc17}. Another detailed study in \cite{Chitimalla2017} claimed that the encapsulation delay would compromise only few kms in fronthaul and it is also possible to reduce jitter, even eliminate it, with some scheduling techniques.  

\subsubsection{Hardware for fronthaul}

The cheapest option for fronthaul is considered to be WDM (Wave Division Multiplexing) over dark fiber \cite{Murphy2015, Fujitsu2014} where the fiber is leased per mile not by the bandwidth. Places where dark fiber is not available, duct sharing or Physical Infrastructure Access (PIA) can also be considered as a solution \cite{Fujitsu2014}, where cables can be placed in access ducts using a license agreement. It is very much clear that countries with fiber rich city infrastructure will benefit far more from these savings than those with low fiber availability.

\begin{figure}
\begin{center}
\includegraphics[scale = 0.7, clip=true]{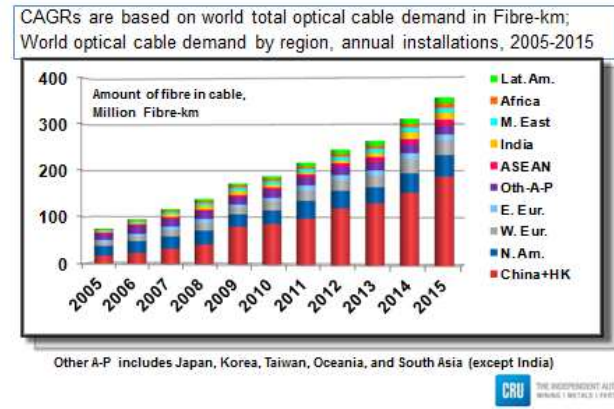}
\caption{Compound annual growth rate (CAGR) for optical fiber around the world.}
\label{fiber}
\end{center}
\end{figure}

Although, fiber deployment is gaining momentum around the world but as shown in Fig.\ \ref{fiber}\footnote{Source: \url{http://wireandcablenews.crugroup.com/wireandcablenews/insights/free/2015/10/4522446/}}, the growth is mainly in China and Hong Kong with Latin America, Africa, Middle East, and South Asia lagging far behind. According to the new statistics released in June 2013, fiber-optic links are used for 15.8 percent of broadband connections in the OECD area with Japan and South Korea leading the list with 68.5\% and 62.8\%\footnote{\url{https://www.cnet.com/uk/news/fast-fiber-optic-broadband-spreads-across-developed-world/}}. Organization for Economic Cooperation and Development (OECD) monitors economic trends in North America, Europe, Australia, Japan, South Korea, Turkey, and some other developed countries and regions. It is very clear that fiber-rich countries will gain the real benefit of the SDN-based and centralized RAN but the region with low economic incentive to deploy optical network will be left behind. It is a serious concern that the advancement in telecommunication technology will further widen the digital divide.

{Other technologies, such as co-axial or twisted pair copper cables, microwave, mmWave, are also explored to provide fronthaul transport network. In {\cite{Artuso2015}}, a mmWave fronthaul is used between small RRUs and macro RRUs. Two different scenarios are tested in preliminary study, i.e., sub-3 GHz and  Eband (71–76, 81–86, and 92–95 GHz). Eband is showed to have approximately 11\% reduced RRUs for the same traffic volume due to availability of broader spectrum.}

\subsubsection{{Software Defined Optical Networks}}

{Another important issue in relevance to optical fronthaul is the developments in software defined optical networks so a complete SDN-based networking solution can be realized. There has been a number of developments regarding software defined optical infrastructure, i.e., transceivers, switches, medium access, performance monitoring, etc. {\cite{Thyagaturu2016-ICST}}. The survey in {\cite{Thyagaturu2016-ICST}} provides a comprehensive account of research in control layer framework, access control, security, QoS monitoring, virtualization mechanisms, etc. Despite the developments, important challenges are still remaining specially in integration of products from different vendors and heterogeneous physical layer technologies, e.g., radio and fibre {\cite{Thyagaturu2016-ICST}}. Security vulnerabilities and scalability issues in centralization of control layer are also needed be explored in further details besides other technical challenges {\cite{Thyagaturu2016-ICST}}.}   

\subsection{Latency of General Purpose Platforms (GPP)}\label{gpp}

{The history of base-station softwarization efforts go back beyond SDN. Potential cost savings and flexibility in implementing and changing protocols were the main drivers behind the efforts. Within SDN ecosystem, base-stations are integral which can have programmable functionalities, i.e., protocols and transmission parameters, either centralized or distributed, implemented on a general purpose computing platform and generic radio heads. By softwarization, the BS and core network functions can be virtualized thus leveraging on the advantages offered by  virtualization such as flexible and agile deployment and management of network functions and agile composition of new network functions. In this context, the performance of GPP became a genuine concern.}

GPPs were not designed for real-time signal processing where the delays should be of the order of tens of microseconds \cite{Tan2011}. Although, GPPs are also improving their speed by finding and eliminating the bottlenecks, but it is yet to be seen if the 1000x capacity and 100x data rate promise of 5G can be supported by a base-station in GPP. Moreover, the promise of saving CAPEX in SDN through GPP may not be as huge as the predictions, as the cost of specialized ASIC hardware is also coming down with time \cite{Fujitsu2014}. The actual cost saving should also consider this downward trend. 

There has been significant efforts in implementing LTE eNodeB in software with promising performance results \cite{Tan2011, Kai.chinacom12,Nikaein.mcs15}. The latency values were shown to be decreasing with processor speed. Special techniques are used in the software to improve latency using parallelism, replacing computations with lookup tables, kernel patches for time guarantees, etc. So far, data rates up to 43.8Mbps on a 20MHz channel are shown to be supported by a LTE base-station in software. It is still a long way to go as far as 5G goals are concerned. In this section, we will discuss prominent efforts in this respect and assess the gaps in the current state of the art.

\cite{Tan2011} presents, Sora, a fully programmable software radio platform on GPP from Microsoft. It was the first effort to program LTE on a commodity general purpose computer. Sora implemented LTE uplink Rx PHY that supports up to 43.8Mbps data rate on a 20MHz channel. Sora, however, uses some hardware components for high throughput and low latency data transfer between radio front-end and host memories \cite{Tan2011}. They bridged the synchronous transfer of high data rate RF digitized signal and asynchronous processing of a GPP through FPGA FIFOs (buffers) and on-board memory with the support of PCIe bus standard for high throughput and low latency. They have also provided a separate path for low latency control data. They have used new software techniques, such as, use of lookup tables instead of computations, use of special kernel service for core dedication in multi-core processor environment, etc., to improve the latency for real-time PHY processing. Sora does provide a good starting point but it does not include softwarization of downlink LTE which is also important and requires more complex FFT processing. Also, realizing the projected data rates for 5G, i.e., in the order of tens of Gbps, it is still a long way to go. Moreover, the real-time processing of CoMP (Coordinated Multi Point) interference management scheme to reap the benefits of centralization would be a big challenge as well on software platforms.
 
{Virtual WiMax base-station pool is developed and evaluated in {\cite{Zhu.cf11}} which can meet system requirements including synchronization, latency and jitter.The implementation uses a FPGA based adapter board between radio head and virtual base-station pool implemented on GPP and can support up to 10MHz signal bandwidth.} 
 
In another effort, reported in  \cite{Kai.chinacom12}, LTE eNodeB prototype is implemented using single core GPP. The prototype was tested for latency in receiving uplink signals and transmitting downlink signals and the results did not meet LTE standards \cite{Kai.chinacom12}. The authors, though, expressed the hope that a multi-core platform would bring the delays within LTE specifications. 

OpenRadio \cite{Bansal.hotsdn12} explored the softwarization of PHY layer of a base-station. OpenRadio uses hardware accelerators for highly computational blocks, such as convolutional coding, Turbo coding, etc. and separates the decision rules from the protocols to be put in the software in order to meet strict deadlines. OpenRadio is also concerned about the software based decision plane not to become very heavy as it would incur prohibitive inter-core communication overheads implying that decision/processing plane separation might not be the best design choice \cite{Bansal.hotsdn12}.

A notable effort in this regard is from EURECOM called OpenAir Interface\footnote{\url{http://www.openairinterface.org/}}. OpenAir Interface (OAI) is an open source implementation of LTE Release 8.6 with parts of Release 10. The maximum expected throughput is reported to be 36Mbps on a 10Mhz downlink channel. Experiments on OpenAir Interface show that the BBU processing delay gets lower with processor speed \cite{Nikaein.mcs15}. A 3GHz single core processor is shown to comply with the HARQ delay requirements of uplink and downlink processing \cite{Nikaein.mcs15}. However, an independent performance study \cite{Yeoh.icact16} shows that the downlink execution time increases with the load and could reach over 200ms for PDCP payload size of 1500bytes. OpenAir Interface is an on-going project, it is yet to see how the base-station software fare for 5G throughputs and data speeds. 

{In a recent effort {\cite{Wang2017}}, both EPC and eNodeB are deployed in one mini Intel commercial PC (i7-5557U). All protocol layers of eNodeB are implemented on the GPP platform with X86 CPU instruction set. The communication between RF block and BBU block is through high speed USB 3.0 interface. The software based eNodeB has USRP B210 software defined radio for transmission and reception. Currently, this implementation can support stable real-time signal transmissions with a bandwidth of 10MHz and a stable data rate of 32Mbps to multiple commercial mobile terminals. The authors also commented about the power consumption of their LTE implementation which is comparatively more than the dedicated hardware.}

srsLTE is an open source implementation of LTE PHY only on GPP \cite{Gomez-Miguelez2016} with extensive use of data and instruction parallelism and use of different kernels of SIMD (Single Instruction Multiple Data) code for different architectures. One processing core showed to have met the delay deadlines of LTE but multiple cores were recommended to be used for processing of the rest of the layers. A commercial version of srsLTE is also available. 

A LTE eNodeB in software is commercialized by Amarisoft\footnote{\url{http://www.amarisoft.com/?p=amarilte}}, where they claim to have a fully LTE release 13 complaint software which can be used with USRP\footnote{\url{https://www.ettus.com/product/category/USRP-Networked-Series}} radio front ends. The software seems to be meeting all the timing requirements but no documentation is available to explain how did they achieve it. The company also claims that 500 UEs can be supported by a single PC but the maximum achievable throughput is not mentioned. 

A proprietary virtualized access (vAccess) development platform from Freescale \cite{Rouwet2015} is created to facilitate the development of SDN/NFV products. It is built over OpenStack cloud computing platform with hardware accelerators for L1 real-time processing and Linux patch to support bounded guarantees for application start time. A LTE smart base-station built over the vAccess platform is also presented in the market\footnote{\url{http://sooktha.com/sbs.html}}.

{In 2016, NEC announced NFV CloudRAN\footnote{\url{https://www.sdxcentral.com/products/nfv-c-ran/}} software which provides flexible functional split between centralized digital unit and radio unit of a CloudRAN. The software application runs on GPP platforms but requires hardware accelerator for L1 processing. Similarly, ASCOS's virtual base-station\footnote{\url{https://www.sdxcentral.com/products/asocs-virtual-base-station-vbs/}}, which came out in 2017, also uses hardware accelerators for L1 signal processing.}

There are other efforts to implement LTE in open-source software for research testbeds. OpenLTE\footnote{\url{https://sourceforge.net/projects/openlte/}} and gr-lte\footnote{\url{https://github.com/kit-cel/gr-lte}} are notable examples but they are not for real-time transmission and reception of LTE signals. gr-lte is only a receiver which is based on GNU radio. 

A reconfigurable and programmable SDR platform is proposed in \cite{Karaz.etsi16} which uses hybrid FPGA to meet all timing requirements instead of GPP. However, it is noted in the paper that reconfiguration of hybrid FPGA is a very slow process and it is not straightforward to do it on the fly.

{\renewcommand{\arraystretch}{1.2}
\begin{center}
\begin{table*}
\centering
\small
\caption{Summary of notable LTE eNodeB software implementations.}\label{LTE-Soft}
\begin{tabular}{ l| l l l}
\hline
 Proposals & In software & Dedicated hardware used or not & Performance\\
\hline\hline 
\multirow{2}{*}{Sora \cite{Tan2011}} & \multirow{2}{*}{Uplink Rx PHY} &  between high data-rate RF & supports upto 43.8Mbps on 20MHz \\ 
	&  &  front-end and host memories & channel \\  \hline
\multirow{2}{*}{\cite{Kai.chinacom12}} & \multirow{2}{*}{Release 8} & \multirow{2}{*}{Same as Sora} & Single core processing did not meet\\
	& & & timing requirements\\ \hline
\multirow{2}{*}{OpenRadio \cite{Bansal.hotsdn12}} & \multirow{2}{*}{PHY} & hardware accelerators & \multirow{2}{*}{No reported results for LTE} \\
	& & for coding blocks& \\\hline
OpenAir \cite{Nikaein.mcs15} & Release 8.6 (part of 10) & None & 36Mbps on 10MHz downlink channel\\\hline
{{\cite{Wang2017}}} & {LTE} & {None} & {supports upto 32Mbps on 10MHz}\\ \hline
srsLTE  \cite{Gomez-Miguelez2016} & PHY & None & Single core met the timing requirements\\\hline
Amarisoft (Commercial) \footnote{\url{http://www.amarisoft.com/?p=amarilte}} & Release 13 & None & supports 500UEs per PC\\\hline
{NEC NFV C-RAN\footnote{\url{https://www.sdxcentral.com/products/nfv-c-ran/}}} & {All layers} & {hardware accelerator} & {Carrier-grade C-RAN}\\\hline
{ASCOS vBase-station\footnote{\url{https://www.sdxcentral.com/products/asocs-virtual-base-station-vbs/}}} & {All layers} & {hardware accelerator}  & {Indoor base-station}\\
\hline\hline
\end{tabular}
\end{table*}
\end{center}
}

\subsubsection{{Lessons Learned}}

A summary of notable software implementation of LTE eNodeB is provided in Table \ref{LTE-Soft}, where it is very clear to see the current state of the art. All of the LTE eNodeB software implementations, open and closed, are relatively recent and unless independent and comprehensive trials are done, it would not be clear that they have resolved the latency issue of GPP or not. Although, they definitely show a lot of promise and potential of the technology. We expect that like every technology, GPPs will also improve and multi-core processors will resolve the bottlenecks slowing down their speed. 

Moreover, as we have observed while discussing fronthaul, that leaving PHY at the RRUs could save the capacity of fronthaul and it is very difficult to imagine that network operators would open such low-layer networking to the third party users. We really do not loose much if PHY, which is also very computationally intensive and latency critical specially in 5G, remains in special hardware at the base-stations or RRUs, at least in the initial phase of SDN roll-out. 

\subsection{Backward Compatibility}

Not long ago, the carrier networks have deployed their 4G LTE systems. The systems are performing very well and getting the most out of them by enhancing their capacity and support for number of users is a straightforward business opportunity. SDN technology comes with big promises but we cannot expect that the carrier operators and owners will decommission the recently-erected network. Most probably, 4G-LTE equipment will remain in function for some number of coming years. In view of this, any new technology should have the essential feature of co-existing and working with the legacy systems. In this section, we will review the work which has been done on the compatibility issues of SDN-based mobile networks and legacy systems. In legacy systems, co-existence with 4G equipment is the most important. 

{The recently approved 3GPP non-standalone 5G standard\footnote{\url{http://www.3gpp.org/news-events/3gpp-news/1929-nsa_nr_5g}} talks about the scenarios where LTE's eNB and 5G NR's (New Radio) gNB (next generation NB) co-exist with EPC and/or NGC (Next Generation Core). The coexistence is possible through evolved base-station eLTE eNB which can connect to EPC as well as NGC and new interfaces, such as, Xx (between LTE eNB, connected to EPC, and gNB) or Xn (between eLTE eNB and gNB with either one or both are connected to NGC) {\cite{3gpp.38.2017}}. NGC is a step towards a fully softwarized core by complete separation of control and user plane and modularizing the management functions. The non-standalone 5G allows an intermediate phase with backward compatibility to LTE.} 

Hybrid SDN network where traditional or legacy nodes co-exist with SDN-enabled nodes have been closely investigated in case of wired and fixed networks and the state of the art is way ahead of respective mobile network research. Firstly, an important point to note here is that OpenFlow protocol, which enables the switches to act as dump data pipes by following a flow table provided by the controller, can also be implemented on commercial Ethernet switches and routers \cite{McKeown2008} making them SDN-enabled. The simplest approach could be the dual stack approach where all network nodes have SDN interface along with an interface for normal processing \cite{McKeown2008}. This approach, however, requires substantial effort to implement the software on each and every switch of the network.

Moreover, there are a number of proposals which has shown different mechanisms to successfully operate hybrid SDN networks with some SDN/OpenFlow switches and many non-SDN or legacy switches \cite{Levin.usenix14, Markovitch.icnp15, Jin2016, Jin.sosr15}. The performance of hybrid SDN networks show real promise and the architecture can actually provide a long-term solution rather than a quick fix for the transitional period. In \cite{Levin.usenix14} a small number of SDN switches, strategically placed, are shown to be sufficient for the network to reap the benefits of SDN and behave like a full SDN network by enforcing SDN policies, e.g., access control. In \cite{Jin2016} and \cite{Jin.sosr15}, the legacy switches are manipulated by special messages to forward the packet over the desired route. 10\% penetration of SDN switches are shown to be sufficient to implement failure detection and route recovery mechanism through SDN controller in \cite{Markovitch.icnp15}. These approaches for managing hybrid networks, however, incur significant management complexity, as they control legacy and SDN switches via different mechanisms \cite{Jin2016}. Moreover, the existing techniques on hybrid SDN networks successfully show implementation of SDN policies of routing, access control, flow isolation, failure recovery, etc., but it is not possible to efficiently virtualize the legacy resources via SDN controller without using additional virtualization mechanisms such as hypervisors etc. 

All the above discussed schemes are designed for fixed networks and not for radio access network where the SDN-based RRUs will need to co-exist with eNodeBs, 2/3G base-stations, and RRUs of CloudRAN
. RRU of CloudRAN may not exactly be the same as RRU of SDN-based mobile network. In the context of mobile networks, the research is way behind than the fixed wired networks and there are only very few attempts to propose solutions for the backward compatibility issue. 


Mobileflow \cite{Pentikousis2013} is one of the earliest solutions to resolve the compatibility issue between SDN-based networks and legacy systems using virtualization techniques. It defines a mobile flow controller which can be used to implement the LTE control plane via network applications. The network applications provide all LTE control plane functionalities, e.g., control for S-GW  (Serving Gateway) and P-GW (PDN or Packet data network Gateway) etc. Forwarding plane is defined through mobileflow forwarding engine which work similar to OpenFlow switches by following the routing rules given by the mobileflow controller. They also have radio interfaces to connect to UEs.  The authors claim that operators can use this architecture in one part of the network to be used with the legacy equipment and then move to a more flat control model, without LTE control plane entities, in future by changing the virtual machines. They have validated the architecture through a prototype. However, the most important part of mobile network, i.e., eNodeB is not decomposed into the control and data plane and it is not shown how they will be mapped into the mobileflow forwarding engine and associated radio interface. In the prototype, an eNodeB was remained as it is and the mobileflow forwarding engine was used as the first node after eNodeB to direct the traffic. Although, not explicitly mentioned, mobileflow is in fact a SDN-based core network alternative to be used in place of 3G/4G core networks with option of moving towards a full SDN network in future through software update on the controller. The details about how to implement EPC through OpenFlow switches are also mentioned in \cite{Kempf2012}. 

In optical fiber domain, \cite{Cvijetic2014} showed SDN-controlled optical topology-reconfigurable mobile fronthaul achieving  10Gb/s peak rates with $<$7$\mu$s back-to-back transmission latency and 29.6dB total power budget while maintaining backward compatibility with legacy fiber deployment.

A proxy based solution for backward compatibility of SDN-based core network is proposed in Softcell \cite{Jin.conext13}. Each base-station has a proxy serving as GTP (Gateway Tunnel Protocol) end-point. Softcell implements LTE core network using commodity switches which then carry normal IP traffic between base-stations and Internet. Softcell, however, does not talk about softwarization of the base-station. 

Heterogeneous CRAN \cite{Peng2014} also claims backward compatibility with legacy cellular systems and other CloudRANs as along with RRUs and centralized BBU pool, it also introduces some high power nodes or base-stations providing control signaling. These high power base-stations are similar to macro base-stations. According to the authors of \cite{Peng2014}, these nodes are critical to guarantee backward compatibility where multiple heterogeneous radio networks converge and this architecture can take advantages of both CloudRAN and cellular networks. 

{Similar ideas were presented in {\cite{Demestichas2013}} under the title of Fusion Net, where macro cells comprise of host layer providing anchoring points to the users. However, the traffic is dynamically transmitted and aggregated from a boosting layer consisting of Wireless LAN (WLAN) access points, small cells, RRUs, etc. The paper also discussed SDN and NFV as emerging technologies but did not bring them in the context of Fusion Net.}

In \cite{Trivisonno2015}, the discussion about backward compatibility leads to the mechanisms where SDN-based network (RAN and core) can be configured to be used as LTE 4G network. This is beneficial if 4G UE (User's Equipment) is being served by the network. However, the UE should have the appropriate applications to instantiate control and data transmissions. Moreover, the system still does not show the co-existence capability with 4G system where seamless handover and resource management could be possible between the two systems. According to \cite{Bernardos2014}, the main property for a SDN enabled backward compatible network is to implement standardized interfaces for interworking with legacy networks.

Another effort of addressing backward compatibility issue is presented in \cite{Kyung2015} where a 4G mobile communication system is shown to be used as a SDN-based mobile network via the separation of Edge Control Plane (ECP) and edge switches from the 4G core network. ECP contains all the intelligence and LTE control plane, i.e., MME, HSS, PCRF, etc., where as the 4G network only performs packet transmission. The access nodes, i.e., eNodeB, remain unchanged. \cite{Kyung2015} also talks about serving 3G and WiFi nodes through similar SDN-fixed 4G network. This scheme is a good proposal for first step towards the full SDN implementation but it requires substantial management rework in 4G networks. It is yet to see if it will make a good business case to revamp the control plane of already erected 4G network or will it be easier to install the new 5G systems, already arriving in the market, with similar SDN fixes as in \cite{Kyung2015}. The proposed SDN-based mobile network architecture in \cite{Costa-Requena.eucnc15} maintains the 3GPP interfaces so the architecture can also be used with legacy network. The architecture moves EPC to cloud but leaves eNodeB unchanged.

\subsubsection{{Lessons Learned}}

In order to summarize the issue and state of the art of the relevant solutions, we believe that interoperability or backward compatibility may take different definitions but coexistence with 4G or legacy networks should include the following scenarios:
\begin{enumerate}
\item A 4G or legacy UE in the coverage of SDN-based mobile network. \cite{Trivisonno2015} tried to look at this type of scenarios where a SDN-based network can be programmed to serve a legacy/4G UE, although, the UE is required to have appropriate application. This scenario is, however, not really interesting from practical point of view since the upgrades in mobile handsets are much more rapid than the enhancements in the access network.
\item A legacy base-station or eNodeB is attached to a SDN-based core network. As shown in \cite{Trivisonno2015}, the SDN-based network can be programmed to serve as a 4G network.  In Softcell \cite{Jin.conext13}, the base-stations are assigned proxies which can act as GTP end-points to communicate with SDN-based core network. The major property identified in the literature for interworking of new and 4G systems is to maintain the standard 4G interfaces and GTP tunneling \cite{Bernardos2014}.  
{In 3GPP's 5G non-standalone architecture, this scenario is similar to option 4/4a/7/7a {\cite{3gpp.38.2017}}. The solution is to enhance eNB and define additional interfaces. The standard also defines the situation when a 5G gNB (next generation NodeB) could be connected to an EPC through assistance from LTE eNB using a new interface Xx.} 
\item eNodeB (eNB) or heterogeneous legacy base-stations or phantom cells and RRUs of SDN-based mobile network share a geographical area
. The SDN network should take the spectrum resources used by the eNodeB or legacy cells into account while distributing resources among its RRUs. According to our search, no such scheme exists in the literature. 
\item Handover between 4G and SDN-based mobile network. Mobility management mechanisms should be aware of the possible handover between the systems. For SDN-based core network, Softcell \cite{Jin.conext13} suggests implementation of S1-MME and S10 interfaces so it can handover to and from legacy LTE EPC. 
{3GPP's recently approved standard for non-standalone 5G NR (New Radio) defines the handover between multiple RAT (Radio Access Technologies). The solution assumed a new interface between EPC and NGC (Next Generation Core) {\cite{3gpp.38.2017}}.}
\end{enumerate}

\subsection{Disruptive Deployment}

The issues of disruptive deployment and backward compatibility have a lot in common and the schemes discussed in the above section are in fact making it possible to deploy SDN networks in an evolutionary fashion. {A good example is the recently approved 5G NR non-standalone standard which is an intermediate step towards full 5G NR deployment and it allows the operators to start experimenting with 5G equipment right away.} However, the work on fixed networks are much more mature and sophisticated than wireless access networks and there are a number of investigations which showed that a small number of SDN switches are sufficient for the network to reap benefits of SDN in the transition phase \cite{Levin.usenix14, Jin2016,Jin.sosr15}, and \cite{Markovitch.icnp15}. 

For the radio access networks, schemes discussed in the previous section and introduced in \cite{Trivisonno2015, Kyung2015} provide the initial approaches towards evolutionary deployment of SDN-based mobile network building operators' confidence before major upgrade of the system. SDCN \cite{Lai2015} talks about a phased approach towards fully programmable centralized RAN. In the first phase, the control or signaling plane of legacy systems, i.e., radio resource scheduling, hand-off, paging, etc., is moved to a logically centralized location or cloud. In the second phase the data planes of the base-stations can be implemented by SDR (Software Defined Radio) on commodity of the shelf (COTS) hardware. In the third phase CloudRAN is implemented with BBU pool using general purpose platforms and RRUs on the remote site. 

Similar step-wise idea is presented in \cite{Costa-Requena.eucnc15}. The proposed SDN-based mobile network architecture in \cite{Costa-Requena.eucnc15} suggests a 3 step migration process from legacy network to fully complaint SDN network. It moves EPC to cloud while maintaining 3GPP interfaces with legacy nodes. In the second step, OpenFlow switch is introduced and GTP tunneling is replaced with MPLS tagging. In the third step, legacy nodes (S/P-GW) are removed from the network. The eNodeB stays the same in this architecture \cite{Kempf2012}.

Softcell \cite{Jin.conext13} proposed a SDN based LTE core network using commodity switches and routers. The major contribution lies in compressing thousands of rules and policies into manageable routing rules for off-the-shelf switches. They proposed that the carrier put a proxy on each base-station which can act as GTP end-point. The Softcell switches then carry normal IP traffic between base-stations and core network. The network can have Softcells in one part while legacy core in the rest in the initial phase and can slowly move to full Softcell implementation. The proposal, however, does not cover the radio side of the network and base-stations are not made programmable.

Mobileflow \cite{Pentikousis2013} also allows part of the core network to be SDN-enabled but still work with legacy networks and base-stations. The architecture supports GTP tunneling and maintains the EPC standardized interfaces for compatibility. Independent implementation studies are required to confirm the claims of Softcell \cite{Jin.conext13} and Mobileflow \cite{Pentikousis2013} that they can co-exist and work with legacy systems. Moreover, it is important to check the performance of such hybrid systems and how easy or difficult it is to implement SDN rules and policies on the flows. 

An interesting classification of proposed SDN-based mobile network architectures based on revolutionary or evolutionary migration is done in \cite{Nguyen2016}. Most of the evolutionary architectures are shown to have preserved GTP tunneling. OpenFlow, in general, does not support GTP tunneling and that is quoted as a major reason to use MPLS labeling \cite{Costa-Requena.eucnc15} in SDN networks. Although, there are attempts to extend OpenFlow to support GTP tunneling \cite{Kempf2012}.  \cite{Bernardos2014} generalizes the evolutionary approach by identifying the importance of maintaining standardized interfaces by SDN network to have interworking with the legacy networks. Similar considerations were part of the evolutionary approach in \cite{Rost2014} to move towards a flexible architecture for RANaaS (RAN as a Service). RANaaS chooses level of centralization according to the requirements
and network state.

Three different alternative are being discussed for SDN deployment: overlay, white-box, or custom hardware\footnote{\url{http://www.enterprisenetworkingplanet.com/netsysm/making-the-business-case-for-sdn-1.html}}. Overlay technologies on existing infrastructure, such as, hypervisor for virtualization will offer minimal disruption but will not provide flexibility to use the hardware efficiently. Some optimization opportunities may be missed. Vendor-specific customized solutions using ASIC can also be deployed with minimal interruptions but will not provide a truly real-time programmable and economical platform. On the other hand, white-box solution requires a lot of standardization across the hardware and software domains to simplify the deployment and systematize the upgrade procedures.

\subsubsection{{Lessons Learned}}

Most of the ideas of evolutionary deployment of SDN-based mobile networks remains at a  high-level and prototype implementation are necessary to ascertain their workability. From business perspective, the point solutions which can work with minimal disruption but can result in quick monetization are the most favorable. Now, the question is: what these point solutions will be in the SDN context filling in the puzzle pieces to complete the bigger picture.

{The idea of non-standalone 5G NR (New Radio) in 3GPP's recently approved standard {\cite{3gpp.38.2017}} is an intermediate step and a good trend to follow. The transitional phase is achieved through additional interface designs and evolving current LTE eNB into eLTE eNB which can connect to 5G core. 5G NR is a step towards full softwarization of the network through separation of control and user plane and modularizing of management functions.} 

\subsection{SDN Specific Security Issues}

The openness promised by SDN also introduces additional security risks and vulnerabilities \cite{Fujitsu2014}. Currently, carriers do not allow third party to meddle with the network protocols, routers, and switches, which results in an ossified network but its management and security is comparatively easier. Users' access to the networking fabric will increase occurrences of intentional and unintentional incidents affecting network operations. Moreover, the logically centralized controller is very critical component of the network. If the controller is compromised, the attacker will have tremendous access over the network \cite{Ahmad2015, Liyanage.ngmast15}. On the other hand, data is not co-located with control plane as in traditional network and the attacker needs to re-direct the data to get hold of it \cite{Ahmad2015}. Moreover, the SDN virtualization abstraction layer integrates various ISP platforms while hiding the specific protocol details which makes security management comprehensive and manageable \cite{Ding2014}. Furthermore, the capability of redirection or filtering of the flows based on packet content comes naturally with SDN and can enhance the network security \cite{He2016}. According to \cite{Dabbagh2015}, global network view, self-healing mechanisms, and the additional control capabilities give SDN great security advantage over traditional networks, though, the controller, forwarding plane and the links between them are exposed to new threats and attacks.

Another important issue with SDN networks is the configuration errors \cite{Scott-Hayward2016}. The network allows users and application to configure and program the network which may result in inconsistent policies and flow rules which could be both intentional or unintentional. It is very important to constantly check the network for inconsistent policies without adding a lot of monitoring overhead and overloading the controller with messages. Scalability and availability of the controller is also an issue 
with logically centralized SDN control plane. This issue is also discussed and looked at in the security related literature. A simple solution is to have distributed control where each controller is responsible for a cluster of switches with some overlap with other controllers to have robustness and resilience against failures \cite{Scott-Hayward2016}. Moreover, SDN are developed over GPP platforms which may have their own vulnerabilities. It is important to harden the platform or it will become potential surface for attackers \cite{Ezefibe.ccece16}.

Similar to other aspects of SDN, the work on security of SDN-based wired networks are more mature than in wireless SDN networks \cite{ Ahmad2015}. Although, the work is equally relevant for SDN-based mobile networks as even in the wireless settings, the SDN controller and links connecting it to the switches are still considered to be the most vulnerable points for attack \cite{Ahmad2015,Liyanage.wowmom14}. The wireless part of the system, however, has some unique issues, such as, user mobility, the co-existence of various generations of cellular systems and WiFi owned by different operators, security of wireless link over shared medium, etc., \cite{He2016}.

Surveys of SDN-based networks' vulnerabilities and related solutions are presented in \cite{Ahmad2015, Scott-Hayward2016, Liyanage.ngmast15, Chen2016}. Survey in \cite{Ahmad2015} identifies possibilities of security threats and attacks in SDN in relevance with the traditional STRIDE (Spoofing, Tampering, Repudiation, Information disclosure, DoS (Denial of Service), Elevation of privilege) threat model and discusses respective work and proposed solutions. According to \cite{Ahmad2015}, DoS (Denial of Service) attacks are expected to be frequent in SDN networks than in the traditional networks as significant amount of data will be exchanged between the central controller and the switches. Spoofing may have less chances of occurrence in dynamic SDN network as it depends on tricking a network services based on obsolete information \cite{Ahmad2015}. We have looked at each SDN layer and interfaces and discussed only the important issues identified in the literature. A comprehensive survey of security related work is outside the scope of present paper and we refer the readers to \cite{Ahmad2015} and \cite{Scott-Hayward2016} (more recent) for thorough survey of security related research in SDN networks. The survey in \cite{Chen2016} is exclusively for mobile networks. 

\subsubsection{Controller}

The logically centralized controller is very critical component of the network. If the controller is compromised, the attacker will have tremendous access over the network \cite{Ahmad2015, Liyanage.ngmast15, He2016, Chen2016}. The idea of standby controller, when one of the controller fails, is discussed in \cite{Ding2014}. There are several proposals for distributed control design for scalability and  reliability \cite{Scott-Hayward2016}, although, the distributed control plane and its interface for interworking are not properly explored. A Byzantine mechanism is also proposed where each network element is managed by multiple controllers. Optimization is used for controller assignment to minimize the number of controllers for a given Byzantine fault tolerance \cite{Scott-Hayward2016}.

A simple DoS or DDoS (Distributed Denial of Service) attack can exhaust resources in the controller and also in the forwarding switches. Authentication for legitimate users can mitigate this issue up to an extent but a compromised application with legal credentials can inject faulty flows causing DoS. A flow-based anomaly detection mechanism can resolve this issue \cite{Ahmad2015, Chen2016}.  \cite{Shin.ccs13} develops a connection migration technique on the data plane to classify successful TCP connections from unsuccessful ones in order to protect control plane from saturation attacks. Data plane can send network information to the control plane asynchronously via actuating triggers. According to \cite{Scott-Hayward2016}, with SDN characteristics of dynamic flow table management and distributed control, the threat of SDN-specific DoS attack can be reduced.

On the other hand, the SDN controller provides unprecedented opportunities to develop adaptive and dynamic counter-measures for security threats. Although, the dynamic security control which can adapt to the changing networking conditions may also pose new challenges unknown to conventional networks \cite{Ahmad2015}. A lot of work has been done in SDN-based firewalls which are mostly implemented as northbound API to the controller. The controller stores and dynamically update the Access Control List (ACL) whereas the traditional firewalls are static and may contain obsolete rules as they are manually entered by network administrator \cite{Ahmad2015}.  In the existing security model, policies are in place to enforce the traffic to physically go through security middle-boxes and firewalls. In SDN, however, the paths are logically chosen by controller and it is the responsibility of the controller to ensure that firewalls etc., are enforced \cite{Ezefibe.ccece16}.

\cite{Scott-Hayward2016} surveys the access control mechanisms for OpenFlow where permission systems are developed to apply minimum privilege to the application instead of the current state where full privileges are open to all applications. Similarly, methods around enforcing the permissions are developed to secure the interface between controller and application along with authentication methods to stop unauthenticated users to access the controller.  A summary of possible security threats, causes and solutions is presented in Table \ref{Sec-SDN}.

Another issue raised in \cite{Kreutz2015} is about the resilience problem of the current controller implementations, such as, Floodlight, OpenDayLight, POX, and Beacon. It is claimed that common application bugs are enough to crash existing controllers. From the security point of view, a simple malicious action, such as, changing the value of a data structure in memory can directly affect the working and reliability of the controller \cite{Kreutz2015}. \cite{Chen2016} and \cite{Ezefibe.ccece16} point out that the packet without any match to the flow table is sent to the controller and an attacker can exploit this vulnerability and overwhelm the controller just by sending data. A possible solution is to design switches with enough processing power to forward maximum expected load and not plenty of packets \cite{Ezefibe.ccece16}. According to \cite{Kreutz2015}, SDN development still need to go a long way to achieve the required security and dependability. 

\subsubsection{Southbound interface}

Dedicated special connection can be provided to improve security of the links between controller and its switches, also called the southbound interface, but it does not eliminate the possibility of compromising the communication \cite{Ahmad2015}. The term southbound may suggest direction of flow from the controller to the switches, although, this interface is bidirectional and we are using the terminology to conform with the SDN research community.

Most of the solutions developed for SDN are designed specifically for OpenFlow \cite{Ahmad2015}, although, the threat vectors are independent of any technology and based on the conceptual layers of basic SDN architecture \cite{Kreutz2015}. An important issue with OpenFlow is that Transport Layer Security/Secure Socket Layer (TLS/SSL) encryption is optional and not enforced \cite{Ahmad2015}. Moreover, TLS/SSL based communication is not strong  enough to protect the control channel from IP based attacks \cite{Liyanage.wowmom14}. Furthermore, the digital SSL certificate used for the communication is self-signed and is not very secure \cite{Ezefibe.ccece16}. If the private key of the certificate is stolen, the attacker can eavesdrop and join the network.

If SSL encryption is not used, it is possible to launch an ARP (Address Resolution Protocol) spoofing attack on southbound interface where an attacker's MAC address is linked to a legitimate IP address. Counter-measures to ARP spoofing attack include packet-level monitoring \cite{Ahmad2015}. Similarly, lack of encryption can also cause Man-in-the-Middle (MiM) attack. MiM can misdirect and drop flows and can scan the network and compromise confidentiality. Alternative encryption methods along with proper auditing and logging methods are suggested to counter MiM and verify non-repudiation \cite{Ahmad2015}.

In a distributed controller environment, where each controller is responsible for a cluster of network switches, the interface between controllers to share important information is also vulnerable similar to the southbound interface and can have similar types of attacks \cite{Ahmad2015, Chen2016}. As far as OpenFlow is concerned, this interface is not properly researched and standardized yet and it is still to see what shape the interface will take.

\subsubsection{Northbound interface and application layer}

 The northbound interface is also exposed to significant threats as the user applications with power to access and define network resources and processes can also abuse the network intentionally or unintentionally \cite{Ahmad2015}. There is a lack of standardization for application access management regarding rules for granting privileges, binding mechanisms, auditing, trust mechanism between network management applications and controller, etc., \cite{Liyanage.ngmast15, He2016, Chen2016}.
 
In order to prevent any malicious application to gain access to the controller and network, some advanced solutions are also proposed besides establishing trust and authenticating the user \cite{Scott-Hayward2016}. These solutions uses techniques, such as, resolving the conflicting rules based on the priority of application, securing the controller core by sandboxing each application, and creating an isolation layer between application and control plane for fault isolation \cite{Scott-Hayward2016}.
 
\subsubsection{Data plane}

Forwarding switches are vulnerable to different types of attacks, such as, DoS, modification or injection of fraudulent flow rules, compromise, etc. A compromised switch can discard, slow down, or deviate network traffic \cite{He2016}. Even worse, it can launch a DoS attack on the controller \cite{He2016}. The survey in \cite{Chen2016} also mentions the possibility of flow table overflow in the switches due to an intelligent attacker sending flow requests with slight difference in parameters. 

In OpenFlow, VAVE (Virtual Address Validation Edge) module verifies the address of external packets with no record in the flow table. The module is extended to include binding validation and a data-set of black and white lists to counter IP spoofing attacks \cite{Ahmad2015}. Random and continuous change of host identities are also proposed as a counter-measure where an attackers scans the network to learn flow patterns \cite{Ahmad2015}. According to \cite{Kreutz2015}, some attacks, such as spoofing, is not specific to SDN but has a much larger impact in the SDN. The authors give an example of spoofing the address of the controller and comment that a smart attack which would last only for few seconds just enough to install malicious flow rules is very detrimental and very hard to detect.

In a virtualized environment, it is also very important to build tools which can verify logical separation between virtual networks. In practical networks, a VM (Virtual Machine) may access some resources of another VM sharing the same physical platform, tampering the network information \cite{Ahmad2015}. Formal verification based methods are being proposed to verify the isolation of traffic between network slices \cite{Scott-Hayward2016}. So far, the research is focused on OpenFlow and should be extended for broader SDN deployments.

Configuration issues where intentional or unintentional programming of the network causes conflict in policies and flow rules are also termed as a major problem specific to programmable SDN networks. Not just applications, but multiple entities with write access can override or misconfigure the related flows and compromise the switch functionality \cite{Scott-Hayward2016}. As new applications and new devices are added into the network, security techniques should be in place to make sure network correctness and operability \cite{Scott-Hayward2016}. According to \cite{Scott-Hayward2016}, the area of configuration issues is one of the widely researched theme in SDN and solutions are proposed for detection of network errors and data plane verification. These proposals can be classified as real-time or non-real-time. In non-real-time solutions, symbolic executions are used to test OpenFlow applications for correctness and not causing the network to reach an inconsistent state. Binary decision diagrams are used to check intra-switch misconfiguration \cite{Scott-Hayward2016}. Flow policies and flow requests are also verified not to by-pass firewalls. These verifications may take several minutes and are not designed for real-time anomaly detection \cite{Scott-Hayward2016}. There are real-time tools as well for verification of flow tables (e.g., identifying routing loops, unavailable paths, etc.), conflict resolution for applications in a firewall, and detection of firewall violations \cite{Scott-Hayward2016}. According to \cite{Scott-Hayward2016}, the real-time tools show the evolution in right direction, although, experimental deployments will be needed to bridge the gap between theoretical models and characteristics of live network. 

The survey \cite{Scott-Hayward2016} also lists the northbound APIs which are designed for conflict resolution in policies. They extract non-overlapping policies from the flow rules before instructing the controller to install the flow rules into the switches. The work, however, should be extended for distributed control environment. Policy management techniques are also proposed for conflict resolutions across multiple modules in different layers and also for comprehensive policy update across all the relevant packets and flows \cite{Scott-Hayward2016}. According to \cite{Scott-Hayward2016}, a concern with these policy conflict resolution is their scalability for larger applications or larger networks as all of these techniques are highly computationally extensive. Another alternative is to maintain a strongly consistent data store, instead of conflict resolution in policies, to maintain network consistency.  
 
A summary of identified attacks and their solutions with relevance to SDN-based networks and specially SDN-based mobile networks is presented in Table \ref{Sec-SDN}. As we have discussed before, most of the work is mainly focused over OpenFlow and its mechanisms. Table \ref{Sec-SDN} is definitely not complete and as the state of the art moves forward, new threats and vulnerabilities will be identified. \cite{Liyanage.ngmast15} presented a similar table but we have added information from other publications into the table and talked about the possible solutions to have clear idea about the current state of the art. Some attacks can be classified differently than Table \ref{Sec-SDN} considering the multiple layers and interfaces it may affect as also noted in \cite{Scott-Hayward2016}.

{\renewcommand{\arraystretch}{1.2}
\begin{center}
\begin{table*}
\centering
\small
\caption{Summary of identified vulnerabilities and solutions.}\label{Sec-SDN}
\begin{tabular}{ l| l | l| l|l}
\hline
 SDN layer/interface & Threats & Causes & Solutions & References\\
\hline \hline
\multirow{3}{*}{Application layer} & Fraudulent access & Lack of access control & Access control 					&\cite{Liyanage.ngmast15, Scott-Hayward2016, Chen2016}\\ \cline{2-5}
	& Accountability & Lack of binding mechanisms & Secure standard for APIs & 								\cite{Liyanage.ngmast15}\\ \cline{2-5}
	& Malicious application & Users' credential/identity theft & Anomaly detection & \cite{He2016, Scott-Hayward2016, Chen2016}\\ 
\hline 
\multirow{4}{*}{Northbound interface} & Malicious access & Limited secure API & Strong access control & 		      			\cite{Liyanage.ngmast15} \\ \cline{2-5}
	& \multirow{2}{*}{Policy manipulation} & Lack of binding mechanisms & \multirow{2}{*}{Secure 				standards for interface} & \cite{Liyanage.ngmast15, Scott-Hayward2016} \\ 				
	& &  for applications& & \\\cline{2-5}
	& Fraudulent rule insertion &Malicious applications & Access control & 									\cite{Liyanage.ngmast15,Scott-Hayward2016}\\
\hline
\multirow{7}{*}{Control Plane} & \multirow{2}{*}{DoS, DDoS } & Flooding to affect & Flow level anomaly & \cite{Liyanage.ngmast15, He2016, Ahmad2015} \\ 
	& & legitimate flows & detection & \cite{Scott-Hayward2016, Chen2016}\\ \cline{2-5}
	& \multirow{2}{*}{Compromise} & Visibility and openness  & Isolation of controller, & \cite{Liyanage.ngmast15, He2016}, \\
	& & of controller&  fail-safe mode for switches & \cite{Scott-Hayward2016}\\ \cline{2-5}
  	& \multirow{2}{*}{Unauthorized access} &\multirow{2}{*}{Lack of access control} & Strong and adaptive & 	\multirow{2}{*}{\cite{Liyanage.ngmast15, Scott-Hayward2016}} \\ 
	& & & access control & \\ \cline{2-5}
  	& Privilege escalation & Tampering of access control & Auditing & \cite{Ahmad2015}\\ 
\hline
\multirow{9}{*}{Southbound interface} & Man in the middle attack, & Optional use 								of TLS/SSL & \multirow{2}{*}{Encryption}& \cite{Liyanage.ngmast15, Ahmad2015}, \\ 		
	& repudiation  & encryption & & \cite{He2016, Scott-Hayward2016}\\ \cline{2-5}
 	& TCP level attack & TLS is susceptible & HIP based IPSec tunneling & 									\cite{Liyanage.wowmom14,Liyanage.ngmast15}\\ \cline{2-5}
 	& ARP spoofing & If SSL (optional) is not used & Packet level monitoring & 						\cite{Ahmad2015}\\\cline{2-5} 
  	& \multirow{6}{*}{Scanning attacks} &   & Encryption, & \multirow{6}{*}{\cite{Ahmad2015}}\\
  	& &  & standards for controller access, & \\ 
  	& & Scanning worms,  & anomaly detection for & \\ 
	& & improper controller access,  & scanning worms & \\ 
  	& & & anonymization,  & \\ 
	& & & time-out randomization & \\ \cline{2-5}
 \hline
\multirow{6}{*}{Data plane} & \multirow{2}{*}{IP spoofing} & \multirow{2}{*}{DNS tampering} & 					Authentication, &\multirow{2}{*}{ \cite{Ahmad2015}} \\ 	
	& & & black/white lists for flow rules& \\\cline{2-5}
 	& \multirow{3}{*}{Tampering} & Illegitimate modification & Encryption & 									\cite{Ahmad2015, Scott-Hayward2016, Chen2016}\\\cline{3-5}
 	& & Improper isolation between  & \multirow{2}{*}{Tools to verify slice isolations} & 						\multirow{2}{*}{\cite{Ahmad2015}} \\
 	& &  virtual networks & &\\ \cline{2-5}
 	& Fraudulent flow rules & Compromised switch & Anomaly detection & \cite{He2016, Scott-Hayward2016, Chen2016} \\ 
\hline
Controller-controller & \multicolumn{4}{c}{Similar to southbound interface} \\
interface& \multicolumn{4}{c}{} \\
\hline \hline
\end{tabular}
\end{table*} 
\end{center}
 }
 
\subsubsection{Issues exclusive to SDN-based mobile networks}

The security analysis and counter-measures discussed above are designed for wired networks only and mobile environment features and cross-domain interoperability is not considered in any of them \cite{Ding2014}. In this section, we highlight the security issues and solutions specifically for cellular and wireless networks. As can be seen in this section, security of mobile networks has not gained a lot of attention in the research community. 

\paragraph{Vulnerabilities}

\cite{He2016, Ding2014} identify four major issues in SDN-based mobile networks besides the threats and attacks identified for the wired part of the network. These issues are:1) mobility between different access technologies using different security protocols, 2) mobility between networks owned by different operators, 3) highly constrained computational capacity and storage, and 4) backward compatibility to older generations of mobile systems. These issues create unique security problems in mobile network domain. 

The survey in \cite{Chen2016} looked exclusively at the security issues of SDN-based mobile networks. The paper classified all attacks in terms of SDN layers as also done in the present paper and extends the scope to mobile terminal and radio access medium. Mobile terminals are in ever increasing danger of Trojans and viruses. The terminals are resource constrained in terms of computability and storage and cannot execute heavy duty intrusion detection modules \cite{Chen2016}. It is important to customize the traditional tools to suit mobile terminals' need of light weight safety procedures. Malicious or legitimate RF interference can cause saturation attack at the access points. Threat of DoS from base-stations to the core network is considered low in \cite{Vassilakis.ictf16} as the traffic from the users to the network or the base-station control traffic is generally low.  \cite{Chen2016} also surveyed the security literature in SDR (Software Defined Radio) area and identified RF interference, mobile terminal malware,  and MAC tempering as some of the detrimental attacks in wireless environment. 

Mobile user's privacy and confidential information  can be leaked in an open backhaul network architecture \cite{Vassilakis.ictf16}. Currently, the traffic is tunneled using IPSec to the core network to secure the mobile terminal's identity from any eavesdropper. Similarly, important control traffic can also be compromised over open backhaul. Another issue raised in \cite{Vassilakis.ictf16} is about a malicious access point or base-station gaining access to the network by spoofing its location using a legitimate base-station location. Counter-measures based on GPS tracking can be developed to mitigate the attack. Moreover, the small base-station/access point/remote radio unit is also vulnerable to physical tempering. 

\paragraph{Secure Architectures}

Some work has been done in the SDN-based mobile network area which is also focused on OpenFlow. 
 \cite{Hampel.infocomwkshps13} adds vertical forwarding extension to OpenFlow to deal with mobility and access control management while forwarding to legacy elements of network. \cite{Namal.sdn4fns13} identifies the lack of secure mobility support which make OpenFlow inapplicable in wireless and mobile systems. OpenFlow cannot handle switch mobility, e.g., in a moving train, and secure change of IP addresses in a fast and dynamic mobile network to maintain secure communication between the controller and the switch. The solution in \cite{Namal.sdn4fns13} is based on HIP (Host Identity Protocol) security method to identify a host either by a host identifier or a host identity tag. A flow control agent is also used in \cite{Namal.sdn4fns13} to update controller of the new location information for location based services.

\cite{Liyanage.wowmom14} proposed an architecture for SDN-based mobile networks using HIP (Host Identity Protocol), IPSec tunneling, and SecGW to provide secured control channel. Although, the evaluation used OpenFlow but the proposed architecture is independent of SDN protocol. Resilience against various attacks, such as, DoS, IP spoofing, eavesdropping, etc., is analyzed. The architecture is shown to be vulnerable to volume based DoS attacks and an access control method or firewall is recommended to be used to prevent such attacks. In a follow-up work a software defined monitoring and data collection scheme is implemented on top of the secure control channel to prevent, detect, and react to the security attacks \cite{Liyanage.ngmast15}.

{\renewcommand{\arraystretch}{1.2}
\begin{center}
\begin{table*}
\centering
\small
\caption{Summary of secure SDN-based mobile network architectures.}\label{Sec-mobileSDN}
\begin{tabular}{ c| l | l}
\hline
\multicolumn{1}{c|}{Proposal}& \multicolumn{1}{c|}{Major security feature(s)} & \multicolumn{1}{c}{General attributes} \\ \hline \hline
\cite{Hampel.infocomwkshps13} & Access control for legacy traffic & OpenFlow extended with vertical forwarding\\
\cite{Namal.sdn4fns13} & HIP (Host Identity Protocol) w/ location update  & OpenFlow with secure mobility support \\
\cite{Liyanage.wowmom14} & HIP, IPSec, \& SecGW for control channel& Independent of OpenFlow, resilience against DoS, IP spoofing, \\
	& & eavesdropping, but not volume based DoS\\
\cite{Liyanage.ngmast15} & Traffic monitoring & Extension of \cite{Liyanage.wowmom14} \\
\cite{Ding2014} & Security layer over control layer for access control & Local agents at wireless APs, high-level design\\
\cite{Guerzoni.5guc14} & 3-level unified control for security, mobility, routing, etc. & High-level idea only\\
\cite{Atat2017} & Exploit channel characteristics & Focused on D2D communication\\
\cite{Yan2015} & NFVI-TP (NFV Infrstructure- Trust Platform) & Framework for security and trust within NFV-MANO\\
\cite{Duan2015} & Authenticated handover & Low-latency and simple authentication via centralized controller\\
\cite{Fu2017} & Authentication, access control, periodic key updates, & 3 OpenFlow GWs for UE, RAN, \& network\\
	& routing conflict analyzer&\\
\hline \hline
\end{tabular}
\end{table*} 
\end{center}
 }

\cite{Ding2014} proposed a SDN-enabled security architecture for mobile networks where a security layer is introduced above the control layer to monitor the application and stop the malicious traffic from entering the radio environment. They proposed local agents at the wireless routers and access points for fast responsiveness. Since SDN hides the complexities of multiple domains and their specific protocols, it is relatively easier to provide comprehensive security through SDN architecture.

\cite{Guerzoni.5guc14} proposed a three-level control plane structure for SDN-based mobile networks to implement a unified security, connection, mobility, and routing mechanism. They have included device controller (for UE), an edge controller, implementing 5G network functions through control applications, and an orchestration controller to coordinate utilization of cloud resources. Two versions of edge controller are identified in the paper, one implemented in the cloud and the other in the mobile device for mission-critical control and out-of-coverage services. Without going into the details of the network functions, the paper describes that security along with radio resource management, mobility, etc.,  is implemented though control applications by the edge controller.

{{\cite{Atat2017}} addressed the computational limitations of small wireless devices by proposing a light-weight security framework for D2D communication. They define spatial transmission region for D2D link that can guarantee a minimum secrecy rate through exploiting the physical characteristics of the wireless channels. Analysis shows that as number of users increases, the achievable transmission capacity also increases due to availability of more D2D pairs.} 

{A framework for security and trust for 5G is proposed in {\cite{Yan2015}} in the context of virtualization and SDN. The authors argue that security mechanism in cloud computing are not sufficient for 5G security challenges and required extensions to provide sufficient security services and functions based on NFVI (NFV infrastructure) in a trustworthy and economical way with cooperation of various trustworthy VNFs (Virtual Network Functions). A security and trust framework is proposed within NFV MANO (Management and Orchestration) architecture with NFVI-TP (NFVI Trust Platform) embedded in NFVI by an authorized party. It ensures NFVI platform layer security by providing a root trusted module, a hardware resource, to ensure every component build upon it is certified as trusted. The framework is not evaluated through implementation though. Moreover, the framework and associated discussion is generic enough for fixed networks and does not address any particular issue regarding mobile networking.}

{The idea of taking benefit of SDN features to provide security services is used in {\cite{Duan2015}}, where a low-latency and simple authentication handover scheme is proposed for 5G dense HetNets. The major benefit comes through the centralized control of SDN which is also assumed to host an authentication handover module. Both access points and user need authentication before accessing the network. The simulation results show low handover delays than the traditional method.}

{In a recent publication {\cite{Fu2017}}, an SDN based 5G architecture is proposed with three different OpenFlow gateways and controllers, one each for the mobile user, for the RAN, and for the network. The network is comprised of heterogeneous technologies. The authors identified potential threats and proposed security mechanisms, i.e., authentication, access control, periodic key updates, routing conflict analyzer, for communications between all types of the architectural entities and showed the proof of their feasibility and strength.

A summary of the secure architectures discussed above is presented in Table} \ref{Sec-mobileSDN}, { where the major security feature (or features) and important architectural attributes are also highlighted. As it be seen from the table  that access control is one of the popular mechanisms to secure northbound interface. Also, more detailed solutions emerged as extensions to OpenFlow and the once claiming to be independent of the physical infrastructure have used OpenFlow for proof of concept. The current state of the art with respect to secure SDN-based mobile networks still needs solutions for broader spectrum of threat models including all types of DoS attacks along with comprehensive evaluation of robustness of the solutions.}

\paragraph{Tools}

\cite{Luo2015} proposed a security assessment scheme to quantify security levels to networks. The scheme is proposed for mobile SDN networks although it is generic enough to be used in other networks. The security level of a network is defined in terms of efforts to reach a target error state or in launching an attack. More efforts required to launch an attack would represent resilience or high security level of the system. The authors want to explore alternative methods for quantification of effort in their future work. 

Security issues in embedded mobile devices are discussed in \cite{Skowyra.hpec13} and a learning intrusion detection system is proposed for OpenFlow networks. Statistically different traffic than the user-defined normal traffic is termed as anomalous. This work is not directly related to cellular systems rather it is designed for embedded mobile devices with strict constraints on transmission power and computational capacity. The idea of intrusion detection system to detect DoS though has application in cellular scenarios. A formal verification method is developed for OpenFlow to make sure systems' correctness and security \cite{skowyra.hicons13}. Malware protection for mobile devices based on OpenFlow protocol is discussed in \cite{Jin.gree13}. The protection system uses real traffic analysis of connection establishment packets inside an OpenFlow controller.

Considering the storage and computational limitations of mobile devices for very heavy security applications, \cite{Hurel2014} proposed outsourcing security to cloud using OpenFlow  virtual switch in the device and OpenFlow controller residing in the cloud along with security manager. This method takes away major computational load from the device, employs variety of security services with fast processing. The communication delay is an important factor affecting performance and so far it is not clear if faster processing overcomes the transmission latency or not \cite{Hurel2014}. 

\subsubsection{Summary of outstanding security Issues}

A comprehensive survey of SDN security solutions have been done in \cite{Ahmad2015} and \cite{Scott-Hayward2016}. However, the solutions are still not mature enough to be used for production deployment. Here we list some of the take-away points from our survey of security literature for SDN networks.

\begin{enumerate}
	\item {\bf Comprehensive evaluation}  Although, a lot of work has been done in SDN security but independent experimental evaluations and comprehensive assessment of solutions are required before the technology is ready for deployment. 
 	\item {\bf Controller security} Controller compromise and malfunction can cause huge damage to the network. DoS or DDoS attacks on the controller are the simplest and most expected attacks in SDN environment. Distributed control plane is an important paradigm for reliable and scalable network control. This paradigm is not explored in detail and the inter-controller interface is not yet defined.
 	\item {\bf Access control} There is still lack of standardization for applications to access the controller. The permission schemes \cite{Scott-Hayward2016} should be thoroughly evaluated for correctness and computational feasibility. 
 	\item {\bf Southbound interface} A dedicated link can connect controller to its switch but appropriate level of encryption is needed to provide secure communication. The monetary implications of dedicated link and feasibility should be studied.
 	\item {\bf Isolation between virtual networks} Real-time tools are needed to verify complete logical isolation between the virtual networks sharing the same physical resources.
 	\item {\bf Error/anomaly detection} Anomaly or error detection is an important element in the network where there is high risk of controller/switch compromise. The techniques should be tested for overhead, detection accuracy and time, and it is also necessary to secure the anomaly detector from security attacks.
 	\item {\bf Policy conflict resolution} The SDN network allows users and application to configure and program the network which may result in inconsistent policies and flow rules. There are real-time tools for policy control resolution but they have scalability issues while dealing with large applications or large networks.
 	\item {\bf Data leakage and modification} It is shown in \cite{Scott-Hayward2016}, that there are no solution till date for data leakage and modification is SDN networks. However, encryption is sought to be a valid solution for SDN networks as in traditional networks but the physical separation between controller and switches requires very important control information to be pass over the communication links. Their security should be comprehensively tested for the available encryption standards. 
 	\item {\bf Solutions conforming to SDN principles} SDN networks are supposed to have multi-vendor interoperability, third party application support, integration of virtualization \cite{Scott-Hayward2016}, and fast and dynamic reconfigurations. Any security solution proposed for SDN should also conform to the norms of the network.
\end{enumerate}

\subsection{Clear and compelling business case}\label{business}

Despite the tremendous work in SDN and towards its deployment, an operator survey conducted in 2015 reveals that the majority thinks that lack of clear compelling business cases for SDN is the major obstacle in its widespread adoption and roll-out. The deployment cost came out as the second major obstacle\footnote{\url{http://www.lightreading.com/carrier-sdn/sdn-architectures/defining-use-cases-and-business-cases-for-sdn/a/d-id/716315}}. 

SDN's early adoption in enterprise networks is on its way in big companies, such as Google, Amazon, Facebook, Microsoft, AT\&T, etc.,\footnote{\url{http://searchsdn.techtarget.com/answer/Whats-the-status-of-SDN-deployments-in-the-enterprise}}. A clear business case is present in all these circumstances, i.e., distributed organizations with many branches can benefit greatly by software and cloud based WAN (Wide Area Network) to provide low-cost Internet services easily and quickly. 

Moreover, there is uncertainty around the benefits in terms of CAPEX and OPEX savings claimed by SDN technology. Virtualization, use of GPP, and resource pooling in cloud computing certainly result in cost savings but it is also discussed that cost reduction does not bring the true value of SDN in the picture, rather it is the faster roll-out capability with overall streamlining of network operations which will provide the economic benefits to operators and network owners. {According to a recent survey\footnote{\url{https://www.sdxcentral.com/articles/news/carriers-5g-plans-rooted-sdnnfv-says-ixia-survey/2017/09/}}, the flexibility and scalability of the network with respect to the demand are the main drivers behind investments in 5G technologies. SDN and NFV are the key enablers of a network that flex with the traffic.}
%
%
The fast-fail approach of SDN will result in quicker roll-out of new services but at this stage, it is unclear what services will be commercially viable and what revenues they will generate \cite{Fujitsu2014}. Revenue is a key factor in recovering deployment costs and the Return On Investment (ROI) for SDN is an important consideration. Although, the faster roll-out of services which can reduce the week-long processes to a few minutes may result in multi-million dollar revenue stream for the carrier for services where time to market is a significant success factor.  

\subsubsection{Virtualization}

An NFV approach can reduce the number of physical servers, but NFV is not dependent on SDN and although SDN supports NFV by providing a flexible connectivity platform, virtualization can be achieved without it \cite{Fujitsu2014}. Service orchestration without any change in infrastructure could generate more benefits than SDN/NFV in the short run and a massive transformation of infrastructure can slow the ramp of savings\footnote{\url{http://blog.cimicorp.com/?p=2716}}. On the other hand, the SDN integrated NFV is able to support a sustainable business model for the long term future whereas the traditional custom stack based virtualization approaches will likely fail \cite{ACGResearch2015}. According to \cite{ACGResearch2015}, TCO (Total Cost of Ownership) of traditional virtualization approaches tracks the growth in requirements whereas a common standardized platform based virtualization, e.g., integrated SDN and NFV, breaks the linkage between cost and requirement and can have 62\% lower TCO than traditional approaches.

A ROI analysis by ACG Research \cite{ACGResearch2015} claims that the break-even point can be achieved in one year when traditional virtualization solutions are gradually phased out in favor of a common and standardized platforms, e.g., SDN, for NFV. This analysis considers network transformation cost including equipment and training and predicts over 350\% ROI in 5 years for common platform based virtualization.

In \cite{Bouras.icumt16}, a techno-economic model is developed for SDN based mobile networks for CAPEX and OPEX savings and TCO (Total Cost of Ownership). The models predicts 68\% CAPEX reduction, 63\% OPEX reduction and 69\% TCO reduction when SDN is adopted for mobile networks in comparison to the traditional networks. For base-station virtualization, \cite{Bouras.icumt16} assumes CloudRAN architecture but did not consider the cost of fiber for fronthaul. Similar reductions are reported by ACG Research report \cite{ACGResearchTCO2015} with 68\% lower CAPEX and 67\% lower OPEX by virtualizing EPC. Again, the prohibitive cost of fronthaul is not considered here as well.

\subsubsection{Use of GPP}

The use of GPP or commodity processors instead of special purpose hardware also reduces the cost significantly \cite{Fujitsu2014}. But this cost saving is based on the view that traditional router vendor equipment is over-priced, although, not everyone in the industry shares this view \cite{Fujitsu2014}. Moreover, considering the overall cost of re-organization of telco companies to suit the SDN model, the savings may not be as great as predicted. Similar concerns are also raised about the OPEX savings. The challenge of operating a network with white-box switches from one vendor controlled by a virtual machine from another vendor may be significant and OPEX may be higher than the current technologies \cite{Fujitsu2014}. 

\subsubsection{Resource pooling in cloud}

Recently, there are attempts to quantify the cost savings using cloud-based networking. An initial effort is made in \cite{Naudts.wksdn12} where the Cisco price list is used to determine the CAPEX of two use cases of mobile networks. The classical networks are compared against a hypothetical SDN-based network with similar configuration and another version of a SDN-based network with virtualization and sharing. It is shown that a SDN mobile network can save 13.81\% CAPEX and with virtualization, the savings can go up to 48.04\%. The paper, however, made certain assumptions, such as, the Cisco switches can be upgraded to have OpenFlow which is not impossible. Although, similar assumptions for base-stations are not true. We still do not have a concept for a SDN-enabled base-station. The costs of fronthaul and backhaul are not considered in the study, and can be significant. Moreover, the study did not consider the associated cost of the disruptive nature of network transformation along with training of the staff  \cite{Naudts.wksdn12}. In a follow-up presentation \cite{Naudts.sdn4fn13}, the author discusses qualitative comparisons for OPEX savings considering the centralized and simpler control of SDN networks, though, concern is expressed regarding the initial rise in OPEX to get the new infrastructure to work.

\cite{Checko.opnetwork13} performs simulations using the OPNET simulator to compare baseband processing resources for CloudRAN with distributed RAN. The paper concludes that CloudRAN needs 4 times less resources than distributed RAN based on the daily load forecasts. The cost of fronthaul is not considered in this paper which could be prohibitive for various operational scenarios. In a follow-up paper, however, the authors provide detailed evaluation of split function fronthaul cost savings in terms of different multiplexing gains \cite{Checko2016}. As expected, the fully centralized solution with all BBU functionalities in the cloud gives the best multiplexing gains but this puts a lot of capacity requirements on the fronthaul and can only be feasible for operators with cheap fronthaul access. For low traffic load, the BBU pool should have higher layer processing and requirements on the fronthaul can be relaxed \cite{Checko2016}. The evaluation work in \cite{Wu.hotnets14}, shows a potential of a 30\% reduced requirement for resources when base-stations opt for centralized processing versus the local processing. In the local processing option, each base-station is provided with the resources to handle peak time traffic. 

The theoretical framework introduced in \cite{Suryaprakash2015} shows 10-15\% savings in capital expenditure per square kilometer when cloud-based RAN is compared against a traditional LTE network. However, the theoretical framework lacks some important information, such as, the cost of base-stations in cloud, and assumes that this cost will be lower than traditional systems. The traffic load, usage pattern, and revenue levels in 2015 from an average Finnish network is chosen as a reference case in \cite{Zhang.netsys15} to compare the CAPEX and OPEX of SDN-based LTE versus regular LTE. It is shown that SDN reduces the network related annual CAPEX by 7.72\% and OPEX by 0.31\% compared to non-SDN LTE. The savings are very small comparing to the annual cost of a MNO (Mobile Network Operator) and do not present a compelling case for SDN. However, if cost savings can be translated through to profit, they would represent a significant rise in profit, which could act as a deployment incentive. 

Another recent study \cite{Asenio.ondm16} tries to find the central location and needed equipment for CloudRAN as a minimization problem from CAPEX and OPEX perspective. The results show that maximum or full centralization yields a minimum CAPEX solution for certain LTE-A configurations (40MHz). But interestingly, lower levels of centralization, i.e., more central locations with small IP/MPLS equipment, also yields up to 18\% savings compared to fully centralization when higher capacity CPRI is available to cater for high rate LTE-A data (100MHz). The main reason behind this result is the price gap between big routers (6.72Tb/s and 48 slots) versus small routers (1.40Tb/s and 10 slots).  Similarly, OPEX savings of up to 7\% is shown compared to fully centralized solution mostly due to much higher power consumption of big routers in comparison to the small routers. The lower level centralization is shown to save 37\% CAPEX and 82\% OPEX when compared against fully distributed approach, i.e., the current cellular model. This analysis assumes the availability of fronthaul and the cost calculations only consider the BBU equipment at the central locations.

The fronthaul factor for the cost is considered in the cost model for CloudRAN developed in \cite{Yeganeh2016}. The paper considers the ratio of cost of fiber per km to the cost of one BBU. When this ratio is greater than unity, it is not beneficial to go for a centralized RAN. The paper also showed better results in case of partial centralized RAN solution when cost is minimized with respect to all the factors affecting it \cite{Yeganeh2016}. 

\begin{figure*}
\begin{center}
\includegraphics[scale = 0.55, clip=true]{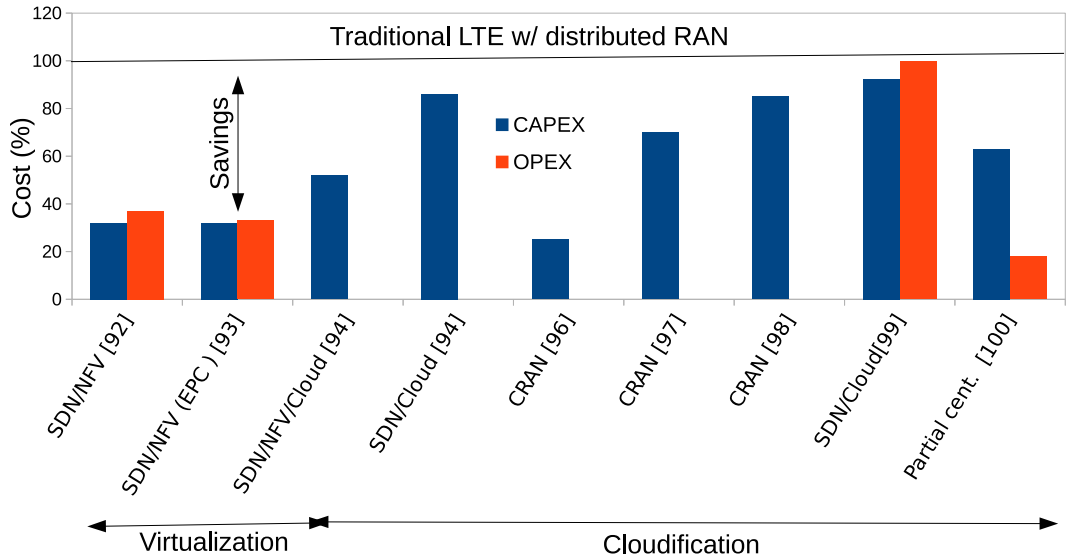}
\caption{CAPEX and OPEX saving estimates when SDN is used with/without virtualization and/or cloudification. CAPEX and OPEX saving estimates for CloudRAN (CRAN) are also plotted. The reference (100\% cost) is traditional LTE network with distributed RAN. Note: details of every estimate method, assumptions, and limitations are discussed in the text.}
\label{SDN-savings}
\vspace{-5mm}
\end{center}
\end{figure*}

{A summary of cost saving estimates are plotted in }Fig.\ \ref{SDN-savings}. {Although, the estimates depend on specific method and assumptions used in the calculations. The numbers should not be compared in strict sense or interpreted as precise benefits. But, it is interesting to see that virtualization seems to have a better potential for cost savings than cloudification. It is not surprising, however, given the potential for end-to-end resource sharing with virtualization. Another factor affecting cloudification savings is the difference in prices of big and small routers {\cite{Asenio.ondm16}} and it may not be cost-effective to replace multiple small routers with one big router. Moreover, an optimal design should also consider the cost of fronthaul.}

\subsubsection{Products and business analysis}

{As the research in SDN-based mobile networks is a step behind wired network developments, there are only a few wireless products emerging in the market whereas there are relatively more WAN-related SDN solutions and products. According to SDxCentral\footnote{\url{https://www.sdxcentral.com}}, the following mobile network related products are announced already:}
\begin{enumerate}
\item {\bf NEC's NFV CloudRAN}\footnote{\url{https://www.sdxcentral.com/products/nfv-c-ran/}} {is a software application which runs on the central digital unit (DU), with COTS servers, but can delegate L1/L2 functionalities to the radio units (RUs) according to the fronthaul condition. It can work with both Ethernet and CPRI as well as one DU can have flexible functional split in different parts of the network.}
\item {\bf ASOCS's virtual Base-station}\footnote{\url{https://www.sdxcentral.com/products/asocs-virtual-base-station-vbs/}} {is a fully virtualized base-station with software of all layers running on GPP servers. It supports flexible baseband partitioning and provides interfaces for digital DAS (Distributed Antenna System). With ease of installation and  integration with any carrier, it offers exceptional total cost of ownership.}
\item {\bf Altiostar’s vRAN solution}\footnote{\url{https://www.sdxcentral.com/products/altiostar-vran-solution/}} {connects the intelligent RRU with the virtualized compute nodes over any transport network. The intelligent RRU understands applications and schedule packets over the air to achieve required QoE (Quality of Experience). This may allow full CloudRAN experience even with constraint fronthaul. Initial trials with SK Telecom have been concluded already where key LTE features are validated. Major customers include SK Telecom, Dali Wireless, etc.  Dali Wireless has recently announced the development of a patented virtual fronthaul interface as an intelligent aggregator-router. That interface makes it possible for multiple operators to connect to multiple BBUs, creating a multipoint-to-multipoint network\footnote{\url{https://www.sdxcentral.com/articles/news/dali-wireless-uses-sdn-to-virtualize-the-fronthaul/2017/10/}}.}
\item {\bf Advantech's Packetarium XLc carrier grade glade server}\footnote{\url{https://www.sdxcentral.com/products/packetarium-xlc-carrier-grade-blade-server-for-virtual-service-edge/}} {is designed for NFV infrastructure for CloudRAN and mobile edge computing (MEC). Its can host 9-12 Intel Xeon processor based blades and supports 720Gbps I/0 and more than 1.2Tbps switching capacity.}
\item {\bf Netsia’s vRAN platform,  called ProGRAN}\footnote{\url{https://www.sdxcentral.com/articles/news/netsias-platform-lets-operators-slice-dice-ran/2017/12/}}, {lets network operators slice the RAN and effectively allocate a portion of the network resources for specific application, such as, safety. Trials to integrate ProGRAN in Telefonica's virtualization project are already underway.}
\item {\bf AT\&T's FlexWare}, {previously called Network on-demand, is based on AT\&T's integrated cloud platform for service orchestration. It uses SDN and NFV to provide software control of different functions for enterprise customers and can be used to offer MEC-capable services. FlexWare is part of AT\&T's 5G trials which will also use mmWave spectrum along with WiFi to provide blanket coverage\footnote{\url{https://www.sdxcentral.com/articles/news/att-will-use-flexware-platform-waco-5g-trial/2017/12/}}.}
\item {\bf BigSwitch's Big Mon} {(Big Monitoring Fabric) can evaluate the traffic of millions of mobile subscribers allowing service providers to monitor their network performance and ensure ultra-high data performance\footnote{\url{https://www.sdxcentral.com/articles/news/big-switchs-big-mon-keeps-eye-mobile-subs/2017/12/}}.}
\end{enumerate}

{There are a number of products and solutions in wired network domain. Some are relevant to mobile networking as well. It is impossible for us to mention all products because of space constraints but some major developments are:}
\begin{enumerate}
\item {\bf Nokia's Network Service Platform (NSP)} {is purpose-built Carrier SDN software for service automation, network optimization and dynamic assurance for delivery of profitable, on-demand network services\footnote{\url{https://networks.nokia.com/products/network-services-platform}}. Integrated with Nokia's Deepfield, NSP can perform real-time analytics for resource optimization and network security.} 
\item {\bf IBM} {has range of products available in SDN family designed to create a unified network architecture which enables cloud and big data analytics and optimizes the entire computing infrastructure, i.e., compute, storage and network resources\footnote{\url{https://www.sdxcentral.com/listings/ibm/}}.}
\item An earlier example is {\bf Juniper's High IQ}, a programmable networking approach, introduced in 2014. ACG Research analyzed three use cases for High IQ from simplifying CPE (Customer Premise Equipment), and pushing the VPN and firewall functions to the cloud, to real-time network self-optimization and elastic traffic. The report indicated compelling business cases through significant CAPEX and TCO savings in all the use cases\footnote{\label{juniper}\url{http://www.juniper.net/us/en/local/pdf/whitepapers/2000581-en.pdf}}. 
\item In 2016, ACG Research's white paper compared {\bf HPE pre-integrated NFV solution} to a bottom-up DIY solution and show the extra OPEX savings of 30\% can be achieved with HPE pre-integrated solution only\footnote{\url{https://www.hpe.com/h20195/V2/Getdocument.aspx?docname=4AA6-4924ENW}}.
\end{enumerate}

\subsubsection{{Lessons Learned}}

From this review there is generally consensus that a positive business case for SDN is evident. A significant issue of network transformation costs is identified, based on the disruption to operational approaches and networking infrastructure. The radical change in the communication network architecture, with full softwarization and centralization, focusing only on the broader CAPEX and OPEX savings in future does not account for transformation costs. A rapid transformation model is very unlikely to be followed, due to the high risk of service disruption. 

A more practical alternative would be an incremental or evolutionary approach to the introduction of SDN into the operational networks. Although, there still need to be individual business drivers or short-term monetization for each incremental implementation to make it appealing for carriers and network owners \cite{Fujitsu2014}. While the evolutionary path to full SDN deployment is also not clear, the individual business cases for the increments are among the big unknowns. Wherever, an enterprise has found some compelling cases to upgrade, the SDN point solutions happen \cite{Fujitsu2014}. Carriers will establish a roadmap of SDN/NFV solutions which will minimise disruption and show short term business benefits \cite{Fujitsu2014}. Thus, to determine the duration of transformation to a full production deployment of SDN is challenging.

\section{Current state of the art of SDN-based mobile network architecture}\label{sdran}


As we have seen in the previous section that there exists a gap between the state where the SDN-based mobile network technology will be ready to be deployed and the current state of the solutions. But not all aspects of the SDN-based mobile network architectural development are covered in the previous section concerning major challenges. In order to really appreciate the current state of the art, we should also look at the major body of relevant literature. A detailed discussion about all the work under SDN-based mobile network architectural development is outside the scope of the current paper. In this section, we have discussed the {major lessons from our study in the form of} common and popular themes and approaches which may become part of the standards.

The earlier papers on SDN-based architectures talk mostly about the benefits as they have to make a case for SDN technology. The architectures they presented are mostly very generic and high-level. As time goes by, the proposed architecture still remains high-level but there are publications which are more focused on the issues and emphasize on details. Recently, there has been a lot of work on specific network services, such as mobility management  \cite{karimzadeh.closer14}, resource management and sharing \cite{Kang2016,Costanzo.icc14,Spapis.iisa14}, CoMP (Coordinated Multi Point) \cite{Bovis.wincom16}, service chaining  \cite{Zhang.icnp13}, network slicing \cite{Katsalis.iccws16} etc. In this paper, we stayed away from network services as a thorough coverage would make this paper too long.  

There has been preliminary implementations with some basic results regarding overheads, cost savings, etc., but a big gap still exists between the current state of the art and the tech-ready state for production deployment. The testbeds available for implementation are also limited by GPP constraints as discussed in Section \ref{gpp} and until we have the platforms suitable for signal processing, it is very difficult to see a real and comprehensive SDN-based mobile network. We have also listed the outstanding issues and problems related to SDN-based mobile network architecture. 

The rest of the section is mainly organized in terms of SDN planes and what has been proposed for each of them and what ideas are repeatedly discussed among the research community. Table \ref{State-SDN} {shows the summarized landscape and an overview of the repeated themes, such as, the realization of EPC nodes as applications in SDN-based mobile network, distributed control plane, and data plane comprising of CloudRAN and heterogeneous radio technologies- e.g., LTE, 3G, and WiFi, etc. for backward compatibility- and OpenFlow (OF) switches. The themes are grouped under the associated SDN layer in Table} \ref{State-SDN}.

{
\renewcommand{\arraystretch}{1.2}
\begin{center}
\begin{table*}
\small
\centering
\caption{Summary of research in SDN-based mobile network architectures.} \label{State-SDN}
\begin{tabular}{l|l|l|l}
\hline
\multicolumn{1}{c|}{SDN Layer}& \multicolumn{1}{c|}{Popular Ideas} & \multicolumn{1}{c|}{References} &\multicolumn{1}{c}{Key Issues/Solutions/Ideas} \\
\hline \hline
 & & \cite{Li.ewsdn12,Pentikousis2013} & 1- S/P-GW split in C/U planes,  \cite{Sama.wiopt14,Nguyen2015,Ksentini.globecom16,3gpp.23.2017} \\
Application & 1- EPC control nodes& \cite{Guerzoni.5guc14,Sama.wiopt14}& 2- S/P-GW CP part is used as application\\
layer & as applications/functions &\cite{Costa-Requena.5guc14,Kabir2014} & 3- CP/UP split is economically optimal \cite{An.netsoft16}. \\
 & & \cite{Nguyen2015,Bradai2015}& 4- Centralized control may create bottlenecks \cite{Basta.sdn4fns13}\\ \hline 
\multirow{14}{*}{Control layer} & \multirow{4}{*}{1- Distributed control} & \cite{Kabir2014,Elgendi.icspcs15} & 1- Delay limit determines cluster size \\ 
 & & \cite{Hamza.imis16,Yang.2016} & 2- Handover between clusters \cite{Hamza.imis16} \\
 & & & 3- Optimal controller placement \cite{Auroux.eucnc15,Ksentini.globecom16}\\
 & & & 4- Local controllers for time-critical tasks \cite{Gudipati.hotsdn13,Ali-Ahmed.sdn4fns13} \\ \cline{2-4}
 & \multirow{2}{*}{2- Hierarchical control} & \cite{Guerzoni.5guc14,Yazici.2014} & 1- Different controller for core, RAN, orchestration, etc.  \\ 
 & & \cite{Trivisonno.iccw15,Vassilakis.icc16} & 2- Fine-grained control difficult for OpenFlow \cite{Yang.2016}\\ \cline{2-4}
 & 3- Global Network View & \cite{Mi.wpmc14,Lu.chinacom15} & 1- Store/update network state information\\
 & (GNV) & \cite{Bradai2015} & 2- Local storage for delay-sensitive data \\ \cline{2-4}
 & \multirow{6}{*}{4- HetNets/multiRAT} & \cite{Yazici.2014,Bernardos2014}& 1- Handover via virtual switches \cite{Chen.iwqos16} and \\
 & & & through virtualization \cite{Riggio.noms14,Vassilakis.icc16}\\
 & & \cite{Sun2015.Network,Chen.iwqos16} & 2- Joint resource allocation \cite{Sun2015.Network,Lai2015,Kang2016}\\
 & & & 3- Traffic offloading \cite{Amani.icc14}\\
 & & & 4- End-to-end QoS \cite{Sun2015.Network}\\
 & & & 5- Energy efficiency \cite{Riggio.noms14}\\ \hline 
\multirow{10}{*}{Data plane} & \multirow{4}{*}{1- CloudRAN} & \cite{Huawei2013, Sabella.fnms13} & 1- Fronthaul capacity and latency\\
 & &  \cite{Riggio.noms14,Wu.hotnets14} & 2- Partial/adaptive centralization \cite{Rost2014, Sabella.fnms13,Jungnickel.icton14,Arnold.eucnc17} \\
 & & \cite{Rost2014,Akyildiz2015}  & 3- Heterogeneous CRAN \cite{Sun2015.Network}\\
 & & \cite{Yang.commag2015,Han2016} & 4- xhaul:backhaul \& fronthaul \cite{Costa-Perez2017}\\ \cline{2-4}
 & 2- Integration of SDR & \cite{Cho2014,Kitindi2017}& \\\cline{2-4}
 & \multirow{3}{*}{3- Exclusively OpenFlow} & \cite{Kempf2012,Li.ewsdn12,Zhang.icnp13,Costanzo.icc14,Akyildiz2015} & 1- Compression of policies \cite{Jin.conext13} \\
 & & \cite{Sama.wiopt14,Mueller.noms14,Kabir2014,Costa-Requena.eucnc15} & 2- Support for (GTP) tunneling \cite{Kempf2012,Heinonen.atc14,Page.icmcis16,Mueller.noms14}\\
 & & \cite{Chourasia.netsoft15,Yang.2016,Page.icmcis16} & 3- Performance requirements/optimization \cite{Li.ewsdn12,Mueller.noms14}\\\cline{2-4}
 & \multirow{2}{*}{4- Control/data split in RAN} & \cite{Liu2014,Artuso2015,Han2016} & 1- Management and virtualization as control modules \cite{Liu2014}\\ 
 & & & 2- Switching-off strategies \cite{Han2016} \\ \hline
 & & \cite{ONF2015,ETSI2015,sdnnfv5GEnablers2017} & 1- Controller be part of OSS/BSS;\\
 & 1- 5 options to & \cite{etsi-mano,Norma2017} & 2- exits as an entity in NFVI; \\
 & integrate SDN & & 3- exists as physical network function (PNF);\\
Integrated & in NFVI & & 4- be instantiated as VNF \cite{Sama2015};\\
virtualization & & & 5- be in VIM (Virtualised Infrastructure Manager) \cite{ONF2015}.\\ \cline{2-4}
 & \multirow{3}{*}{2- Integration in SDN} & \cite{Li.ewsdn12,Liu2014}& 1- Role of hypervisors \cite{Li.ewsdn12,Yousaf.icc14,Akyildiz2015,Haleplidis.ict16} \\
 & & \cite{Costa-Requena.eucnc15,Akyildiz2015} & 2- Virtualization layer above infrastructure \cite{Li.ewsdn12,Ziegler.iccws16,Vassilakis.icc16} \\
 & & \cite{Vassilakis.icc16,Kitindi2017} & 3- Virtualization services as control plane modules \cite{Liu2014}\\
\hline
 & \multirow{3}{*}{1- Resource sharing} & \cite{Spapis.iisa14,Costanzo.icc14,Thyagaturu2016}& 1- Co-primary spectrum sharing \cite{Spapis.iisa14}\\
 & & & 2- Via virtualization \cite{Costanzo.icc14}\\
 & & & 3- Via Smart gateways \cite{Thyagaturu2016}\\ \cline{2-4}
 & \multirow{2}{*}{2- Mobility management} & \cite{karimzadeh.closer14,Valtulina.gcwkshps14,Kuklinski.gcwkshps14} & 1- DMM implementation via X2 \cite{karimzadeh.closer14,Valtulina.gcwkshps14}\\
 SDN & & & 2- Via global network view \cite{Kuklinski.gcwkshps14}\\ \cline{2-4}
 applications & \multirow{4}{*}{3- Multi RAT scenarios} & \cite{Amani.icc14,Cao2017,Hurtado-Borras.icc15} & 1- Traffic offloading \cite{Amani.icc14}\\
 & &\cite{Santos.wowmom16} & 2- End2end routing \cite{Cao2017}\\
 & & & 3- SDN-based backhaul \cite{Hurtado-Borras.icc15} \\
 & & & 4- On-demand small cells \cite{Santos.wowmom16}\\
\hline \hline
\end{tabular}
\end{table*}
\end{center}
}

\subsection{Application Plane}

While transforming LTE architecture into SDN's 3 plane model, EPC control nodes, e.g., MME, HSS, and PCRF are usually suggested to be implemented as application plane modules as shown in Table \ref{State-SDN} \cite{Li.ewsdn12,Pentikousis2013,Guerzoni.5guc14,Sama.wiopt14,Nguyen2015,Costa-Requena.5guc14,Kabir2014,Bradai2015}. 
{Similar ideas are evident into the evolving 5G system's architecture from 3GPP {\cite{3gpp.23.2017}}, where mobility management and policy control functions are connected to other new functions, such as, authentication, application management, etc. via a common message bus. The architecture is a step closer to the softwarization or cloudification of core network.}

However, the case with S/P-GWs is not very clear. 
{In the 5G system architecture from 3GPP, the User Plane Function (UPF) to handle the user plane path of PDU sessions remains in data plane {\cite{3gpp.23.2017}}. UPF in 5G architecture is supposed to the equivalent of PGW-U (User plane part of PGW) in EPC. The equivalent of PGW-C (Control plane part of PGW) is SMF (Session Management Function) in 5G NGC (New Generation Core) {\cite{3gpp.23.2017}}. The separation of control and user plane in 5G NGC is more comprehensive than in EPC. Although, this separation is not exactly the same as the separation of control logic and forwarding hardware defined in SDN.}

A detailed study about which functions of LTE EPC should be moved to cloud in an SDN based system is done in \cite{Basta.sdn4fns13}. They pointed out that as functional blocks move to the cloud, the cost of S/P-GWs goes down as the gateway hardware becomes simpler but data overhead and end-to-end latency increase. The straight forward solution of moving all control functionalities to the cloud may result in unacceptable latency for some applications. Also, cloud infrastructure performance will be a critical factor for high frequency of control plane operations taking place at S/P-GWs. 

On the other hand, moving signaling control and resource management logic to the cloud would allow S/P-GWs to be deployed on distributed elements but would also create bottleneck at the centralized control when forwarding rules are exchanged. Moving resource management back to the data plane would make it more independent and resilient but it would not be possible to take the advantage of centralization through optimal resource allocation. The authors also suggested a hybrid approach where the scenario and traffic requirements would dictate the design but this approach would require state synchronization and orchestration between cloud and data plane to avoid redundant assignments.

A comparative cost analysis of SDN-based EPC and software only EPC, where LTE control and user (C/U) plane are both implemented in cloud, is presented in \cite{An.netsoft16}. The key finding of this study is the realization that a pure software solution where C/U planes are in the cloud is much more expensive and it is economical to leave the user plane (S/P-GWs) outside the cloud and perhaps on the dedicated hardware. An optimal solution for placing SGW-C- i.e.- the controller part of S-GW,  in the network is presented in \cite{Ksentini.globecom16} using game theoretic approaches. The algorithm finds a trade-off between load reduction on SGW-C and reducing relocation of S-GW which is costly for the mobile operators.  

In \cite{Sama.wiopt14}, SGW-C, i.e., the control part of S-GW is used as an application along with MME module and PGW-C (control part of P-GW) with an OpenFlow controller. The data plane, or SGW-D, is comprised of advanced OpenFlow based switches with GTP support. The numerical evaluation show reduction in signaling load with OpenFlow implementation of EPC than the original 3GPP EPC. This work was extended in \cite{Nguyen2015}, where all functional blocks of EPC including S/P-GW control part are implemented as application over mobile controller. The signaling load in full implementation of EPC is shown to be lower than that of \cite{Sama.wiopt14}.
 
\subsection{Control Plane}

One of the important issues in SDN-based networks is controller scalability. Clustering or distributed control plane is explored as a potential solution, although, many important questions, such as, inter-controller coordination, etc., remain unanswered. Hierarchical control planes are also proposed to cater for the time-sensitive control processes on one hand and the need for centralized view for optimal operations on the other. 

\subsubsection{Hierarchical/Distributed Control Plane}

Different proposals are put forward which define domains or clusters for each controller to resolve the scalability issue \cite{Kabir2014,Yazici.2014,Guerzoni.5guc14,Trivisonno.iccw15,Elgendi.icspcs15,Hamza.imis16,Yang.2016,Vassilakis.icc16}. The controllers can be connected to each other in a distributed manner or there could be a higher level controller to coordinate between the lower level controllers. The size of the domain or clusters can be chosen to satisfy the delay constraints.  Hierarchical control plane proposals suggest 2 or 3-tier architectures for scalability \cite{Kabir2014}, ease in management specially in heterogeneous networks \cite{Yazici.2014}, and for fulfilling delay requirements  \cite{Yazici.2014,Elgendi.icspcs15}.

A distributed controller architecture is presented in \cite{Hamza.imis16}, where controller-domains are defined as the eNodeBs served by a particular controller. Handover processes are also defined between eNodeBs in a controller domain and within different controller domains. The handover processes and packet sizes are also revised from 3GPP standards to reduce the signaling overhead. The proof-of-concept, i.e., testbed implementation or simulation experiments are missing from this study to ascertain that the design is complete and adequate for 5G capacity and speeds. The controller domain idea makes the SDN control plane scalable but important details about controller placement and optimization are also not available in this paper. Optimal controller placement is also an open issue. Solutions are presented in \cite{Auroux.eucnc15,Ksentini.globecom16} using optimization function and game theoretic approaches. 

SoftRAN \cite{Gudipati.hotsdn13} also observes that it is cost-effective to leave data plane functionality to the base-stations along with some part of control plane for delay-sensitive decisions, but stressed on the coordination of closely-deployed BS in a dense network though a centralized controller such that the dense base-stations in an area can be viewed as virtual big base-station (big BS). 

Similar ideas of a regional controller which possibly resides in cloud and local controller, located in a macro cell, have been used in \cite{Ali-Ahmed.sdn4fns13}. The SDN-based mobile network model for dense deployments  \cite{Ali-Ahmed.sdn4fns13} is an outcome of FP7 project CROWD with the focus on energy optimization. The paper also proposed a mobility management scheme for their model.

The architectures presented in \cite{Yazici.2014,Vassilakis.icc16} defines a local SDN controller (LSC) for scalability which are responsible for heterogeneous wireless networks in an area and are connected to core SDN controllers (CSC). The paper \cite{Vassilakis.icc16} goes into the details of LSC and what functions it is supposed to perform including content caching, resource and mobility management functions. The authors also showed the backhaul traffic reduction with content caching via simulation experiments. The virtualization layer lies on top of physical infrastructure and each VM (Virtual Machine) communicate to LSC via a local SDN agent. The implementation results are still preliminary and does not show the interworking of different LSCs and does not include offloading mechanisms over heterogeneous technologies. 

\cite{Guerzoni.5guc14} proposed a three-level control plane structure for SDN-based mobile network. They have included device controller (for UE), an edge controller, implementing 5G complaint network functions through control applications, and an orchestration controller to coordinate utilization of cloud resources. Two versions of edge controller are identified in the paper, one implemented in the cloud and the other in the mobile device for mission-critical control and out-of-coverage services. The paper does not go in the detail of physical structure or fronthaul or backhaul issues rather points out the requirements for the control plane architecture. Following up from \cite{Guerzoni.5guc14}, the authors in \cite{Trivisonno.iccw15} define the initial attachment and service requests procedure for the 3-level control plane structure. They show that reconfigurability of the network to use appropriate cloud resources can improve the latency up to 75\% compared to 3GPP Release 12 complaint 4G systems. 

\cite{Yang.2016} describes a high-level architecture for mobile networks with one controller each for core network and RAN. They use a coordinator to provide end-to-end services through both controllers. Some services, such as, firewall, encryption, etc., are implemented through middle-boxes and the controllers decide the flows traversing through these middle-boxes. The paper identifies major challenges of this approach, such as, providing fine-grained control which is not yet possible with OpenFlow. Moreover, the paper provides preliminary performance results through simulations.
 
\subsubsection{Global network view (GNV)}

The paper \cite{Mi.wpmc14} introduces the concept of global network view (GNV) which stores information about network state and can be accessed by network services. The storage can be local for delay-sensitive information, such as, channel state information. The rest of the information regarding flows and load can be stored in a common place. The protocol stack in the architecture of \cite{Mi.wpmc14} is in the modular form which can be orchestrated by the control plane for implementation of specific function, e.g., eNodeB or S-GW. 

The use of GNV with locally and globally stored information is also discussed in \cite{Lu.chinacom15} but only for the context of access network. A similar idea is introduced in \cite{Bradai2015}, where an additional plane called 'knowledge plane' is added on top of the application plane in SDN-based mobile network architecture. This knowledge plane is responsible for providing a global view of network usage which can be used to device new applications to optimize resource utilization. The useful information in the knowledge plane consists of network usage, traffic load, congestion information, etc. 

\subsubsection{HetNets}

The current mobile network environment, in general, consists of heterogeneous radio technologies, such as, 2G/3G/4G and WLAN. Most likely, the future operating scenarios will also have similar characteristics. SDN has the potential of exploiting the different radio resources in a comprehensive manner as the details of the technology are hidden under the abstraction layer and it is much easier to optimize the utilization of available resources in a unified way \cite{Yazici.2014}. 
{Although, the mobility management, joint resource allocation, end-to-end QoS (Quality of Service) assurance, and security issues pose additional challenges specially when UE moves between networks of different ISPs using heterogeneous technologies {\cite{Sun2015.Network}}.} 

Special network selection switches are proposed to achieve handover between different technologies \cite{Chen.iwqos16} and virtualization is also used as a mean to connect different networks to a controller \cite{Riggio.noms14,Vassilakis.icc16}. Traffic offloading schemes are also coming up recently  \cite{Amani.icc14}. In \cite{Kang2016}, a resource allocation mechanism with holistic view is introduced for SDN-based heterogeneous networks. Most of the work in this respect remains at conceptual level and more exploration is needed  \cite{Bernardos2014,Sun2015,Lai2015}. This is one area where optimal resource utilization solution can provide immediate monetization opportunity. 

In \cite{Bernardos2014}, a high-level heterogeneous network architecture is presented with a SDN controller providing network control functionalities to RAN, transport network, service providers, virtual operators, etc. The paper did not go into the details of implementations or low-level architecture of the controller but discusses high-level requirements from different interfaces. The discussion in \cite{Sun2015} also remains at high-level in describing the scenarios and benefits of using SDN to control and operate a HetNet. 

The approach discussed in \cite{Lai2015} proposes a SDN controller cloud providing control to data plane including heterogeneous wireless access systems and BBU cloud for CloudRAN. The SDN controller will provide control to the EPC cloud implemented in GPP as well as physical switches for the core network. The paper provides details about different components of data plane and how they can be specifically controlled with associated issues and benefits. However, the paper still remains a high-level conceptual contribution. 

HetSDN \cite{Chen.iwqos16} also proposed seamless transfer between heterogeneous RANs via virtual switch. They introduced two new components, i.e., SDN-NS (SDN Network Selection) and SDN-HA (SDN Home Agent). SDN-NS uses a virtual switch to seamlessly direct packets to the most suitable wireless interface and SDN-HA is used as the anchor for all traffic for the UE. Another idea to provide seamless movement between LTE and WiFi is through SDA (Software Defined Access) \cite{Sagar2016}. In this architecture, a separate controller, i.e., SDA controller is added on the control plane which has a client side SDA-u on the UE. The also have another element SDA-g (SDA gateway) on the data path which is used as an anchor. Prototype implementation shows proof of design and concept.

VCell \cite{Riggio.noms14} describes an architecture with heterogeneous cells, i.e., macro, micro, femto, etc., where all resources are virtualized in a logically centralized pool. The resources are allocated to the user from a 3D grid of time, space, and frequency resource blocks. The SDN controller assigns resources and manages interference. The main contribution of this paper is to consider heterogeneous cell types while building up the resource grid. The issues with centralization, such as, fronthaul latency and capacity, processing latency of
commodity hardware, etc., are not discussed in \cite{Riggio.noms14}.

{Heterogeneous CloudRANs are also studied in literature where small RRUs and macro RRUs are distinguished by their service area {\cite{Han2016}}. These networks may also contain small or macro base-stations based on the availability of resources and chosen functional split. The macro base-station or macro RRU is shown to provide signaling for the whole service area and small base-stations and RRUs can only be used on-demand optimizing energy efficiency while satisfying QoS constraints.}

\subsection{Data Plane}

CloudRAN seems to be preferred choice for SDN-based mobile network layout, although, it is not always beneficial to centralize all the processing in the cloud \cite{Riggio.noms14,Wu.hotnets14}. The case for partial centralization of processing units is discussed under various themes, such as, in RANaaS (RAN as a Service) the level of centralization is chosen according to the need and network state \cite{Rost2014, Sabella.fnms13}. A plastic or adaptive architecture which provides various levels of centralization according to the operational requirements  \cite{Jungnickel.icton14} could be very desirable from business point of view as it provide further optimization of cost but the technical issues involved in realizing such architecture are not clear at this point. \cite{Arnold.eucnc17} discusses the partial centralization option in their architecture to cater for the fronthaul and backhaul constraints.

{{\cite{Sun2015.Network}} pointed out that, in a CloudRAN integrated HetNet, if multiple standards are being used in the same spectrum, the RRUs can support them only partially. Whenever the standard changes, the BBU is forced to restart instead of sharing multi-standards resource directly.}

\cite{Costa-Perez2017} suggests an architecture for integrating existing and new fronthaul and backhaul networks into a flexible unified 5G 
transport solution using SDN/NFV-based management and orchestration 
(MANO) technology. The paper explores the resource management functional blocks and discusses some use cases.

{Issues and advantages related to the integration of SDR (Software Defined Radio) with SDN are also explored in the literature {\cite{Cho2014}}, {\cite{Kitindi2017}}. Preliminary results have shown that the integrated network with the potential of using spectrum holes and white spaces performs significantly well than the network without SDR {\cite{Cho2014}}.}

\subsubsection{Exclusively OpenFlow based}

OpenFlow seems to be the choice of the researchers at least for now \cite{Kempf2012,Li.ewsdn12,Zhang.icnp13,Costanzo.icc14,Sama.wiopt14,Mueller.noms14,Kabir2014,Costa-Requena.eucnc15,Akyildiz2015,Chourasia.netsoft15,Yang.2016,Page.icmcis16}. OpenFlow and SDN may even appear to be synonyms in some places. During our study, we only came across one paper \cite{Haleplidis.ict16} which advocated the use of ForCES (IETF Forwarding and Control Element Separation framework) instead of OpenFlow. ForCES includes abstraction model for both SDN and NFV as described in RFC 5812 \cite{Haleplidis.ict16}. The major issues in using OpenFlow for the data plane in a cellular network are the lack of GTP tunneling support and the transformation of complex billing and management policies into simple and scalable routing rules for OpenFlow switches.

Among the earliest work on SDN-based mobile network, OpenRoads or OpenFlow Wireless presents a prototype where wireless APs (WiFi and WiMax) are augmented with OpenFlow  \cite{Yap.sigcomm10}. A centralized controller, Network OS (NOX), is used to host the application plug-ins and translate the policies into routing rules for OpenFlow. Applications of mobility management were executed over the prototype. Authors suggested similar OpenFlow additions for LTE, although, LTE base-station and core network is way more complex than a WiFi or WiMax access point and it is not straightforward to translate management and billing policies into simple routing rules \cite{Jin.conext13}. Softcell \cite{Jin.conext13} remarked that the policies of a cellular network translates into thousands of rules and it is important to come up with some logical aggregation techniques so the rules can be fit into today's routers and switches.

The design in \cite{Chourasia.netsoft15} put all the EPC control functions in controller which translates the flow rules for OpenFlow switches used in place of S-GW. P-GW is eliminated from the design. The author claims efficient mobility support with the proposed EPC architecture than the legacy one. The paper overlooked the real problem of scalability and translation of huge lists of policies into simple flow rules for the switch \cite{Jin.conext13}. Until both problem remain unsolved, it would be difficult to realize the architecture.

\cite{Kempf2012} added relevant extensions in the OpenFlow protocol for GTP tunneling and encapsulation. In \cite{Heinonen.atc14}, a new dynamic GPRS tunneling protocol is presented and evaluated for SDN based core network. The protocol allows elastic use of data plane resources and enable cloud to provide on-demand packet processing. Terminating GTP tunnel at the cloud would incur additional delays than terminating it on the fast data path. The paper is preliminary work and more evaluation is needed to understand the scalability and performance with real networks.

The integration of OpenFlow in EPC is discussed in \cite{Page.icmcis16}, where the S-GW is replaced with OpenFlow switches and detailed steps are discussed to keep GTP tunneling and vertical protocol stack for backward compatibility. The paper later introduced different processes like creating, modifying, and deleting a bearer or a session without GTP tunneling for lightweight architecture.

In \cite{Mueller.noms14}, OpenFlow 1.4.0 is extended and optimized for 3GPP EPC including the design of controller, switch, and protocol. The paper first highlights the requirements including the clear definition of northbound APIs, GTP support, latency, etc. The switch is evaluated within OpenEPC implementation for proof-of-concept and scalability.

CellSDN \cite{Li.ewsdn12} identifies additional requirements from OpenFlow switches to provide network services, such as, deep packet inspection, header compression, etc. They emphasized on reducing the controller's load by means of local control agents at the base-station and switches. The interfaces between controller, local agents, and data plane are also defined in terms of requirements and available protocols.  

\subsubsection{Control signaling and data split}

The approach called CONCERT \cite{Liu2014} separates control signaling from data transmission as defined in BCG2 architecture and Phantom cell concept (cf.\ Section \ref{lte-background}). It uses SDN switches with all radio interfaces, computational resources, and servers. These switches are then controlled by the control plane entity called a conductor. All management and virtualization services are defined as control plane modules. Application scenarios are highlighted to focus on the benefits of SDN-based architectures. Technical issues are also briefly explored. 

{The authors in {\cite{Artuso2015}} also touched upon the control-data split in their deployment scenarios but did not go into the details of implementation. {\cite{Han2016}} also discussed the idea in relation to heterogeneous CloudRANs, where macro base-station or RRU can provide signaling to the area and small base-stations or RRUs can only be used on-demand. The switching-off strategies are also discussed in {\cite{Han2016}}.}

\subsection{Integrated Virtualization}

\subsubsection{SDN from NFV Perspective}

\begin{figure}
\begin{center}
\includegraphics[scale = 0.34, clip=true]{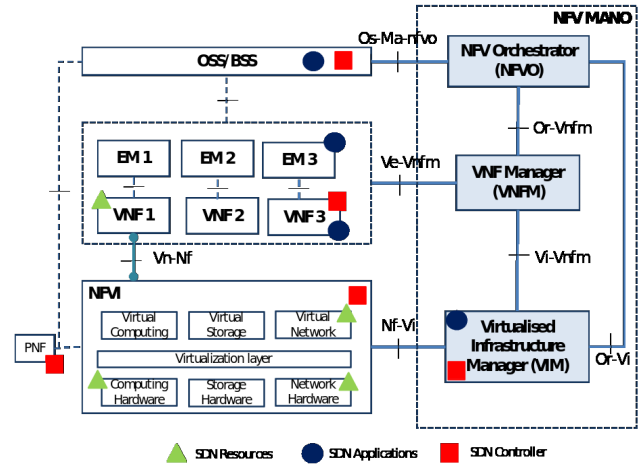}
\caption{Possible options of positioning SDN Resources, SDN Controller and SDN Applications in NFV Architectural Framework \cite{sdnnfv5GEnablers2017}.}
\label{SDNinNFV}
\end{center}
\end{figure}
As mentioned earlier SDN and NFV are complimentary technology enablers for 5G and \cite{sdnnfv5GEnablers2017} provides an overview of both these technologies in the context of 5G network requirements and architecture. From NFV perspective there are different options for deploying SDN within the NFV infrastructure and the various deployment options have been analyzed in detail by ETSI NFV in \cite{ETSI2015}. Figure \ref{SDNinNFV} \cite{sdnnfv5GEnablers2017} summarizes the possible options of integrating SDN application, SDN resources and SDN controller with different entities within the NFV MANO and NFV architecture. There are five integration options for SDN controller to either (i) be part of OSS/BSS (Operations/Business Support Systems), (ii) exist as an entity within NFVI (NFV Infrastructure), (iii) exist as a PNF (Physical Network Function), (iv) be instantiated as a VNF (Virtual Network Function), or (v) be integrated within the VIM (Virtualised Infrastructure Manager). The latter approach, for example, is supported by the ONF SDN architecture \cite{ONF2015} and is also an adopted approach by open source OPNFV platform\footnote{\url{https://www.opnfv.org}}. OPNFV prescribes the integration of SDN controllers like ODL (OpenDayLight) and ONOS (Open Network Operating System) with OpenStack VMM (Virtual Machine Manager). It should be noted that from ETSI NFV perspective, OpenStack platform for all practical purposes is considered as a VIM platform. 

There are also some prominent research projects like 5G NORMA\footnote{\url{https://5gnorma.5g-ppp.eu/}} that leverages on the SDN and NFV concepts in order to develop a novel mobile network architecture that shall provide the necessary adaptability in a resource efficient way able to handle fluctuations in traffic demand resulting from heterogeneous and dynamically changing service portfolios and to changing local context. 

\subsubsection{NFV in SDN Domain}

As the SDN architectural proposals are concerned, NFV integrated SDN architectures make more sense from business as well as technical points of view \cite{Li.ewsdn12,Liu2014,Costa-Requena.eucnc15,Akyildiz2015,Vassilakis.icc16}, {{\cite{Kitindi2017}}}. Designs may include the role of hypervisors \cite{Li.ewsdn12,Yousaf.icc14,Akyildiz2015,Haleplidis.ict16} or virtualization layer above the infrastructure or forwarding plane \cite{Li.ewsdn12,Ziegler.iccws16,Vassilakis.icc16}. In CONCERT \cite{Liu2014}, the virtualization and management services are provided as control plane modules.  These architectures are high-level designs and the relevant issues are not fully explored.

One of the initial attempts to combine NFV and SDN technologies together for mobile EPC is given in \cite{Sama2015}. It shows the complexities and issues involved in achieving the virtualization capacity in a softwarized core network. Firstly, the authors show the use of OpenFlow switch network to replace S/P-GWs of EPC. Important questions left unanswered here, such as, how the huge lists of policies in PCRF can be translated into manageable flow rules for the OpenFlow switches. Secondly, the combined NFV and SDN architecture is proposed which shows two different instantiations of the control plane, i.e. one for NFV and the another for end-to-end connectivity. The two instantiations are connected to each other and also to the NFV management and orchestration unit, at supposedly higher control level. The EPC control plane is also mapped as VNF (Virtual Network Functions) in the NFV infrastructure. Many issues are unanswered regarding the design. Most importantly, how these VNFs will interact with SDN controller to maintain connectivity. Scalability is another issue which needs to be tackled in a centralized EPC architecture. A layered approach providing clear demarcation between actions needed for connectivity and providing services is probably a better way to understand the problem instead of modeling a broad set of NFV and SDN aspect into one flat design as also mentioned in \cite{Sama2015}.

SoftAir \cite{Akyildiz2015} proposed an SDN-based mobile network architecture with network hypervisor to orchestrate virtualization. They have proposed wireless hypervisor and switch hypervisor for SDN enabled base-station and SDN enabled switches. They have designed SDN enabled base-station such that all processing stays at the cloud except modulation/demodulation which are located with antennas at RRUs. The split would reduce the load on CPRI supported fronthaul. 

The paper \cite{Haleplidis.ict16} advocates the use of ForCES (IETF Forwarding and Control Element Separation framework) instead of OpenFlow. The authors build a proof-of-concept focusing on GTP tunneling. The proof-of-concept involved a hypervisor which instantiate VMs and deploy necessary LFB (Logical Forwarding Block), an element of the forwarding plane in ForCES. They build an application to provision and destroy GTP tunnels and collect statistics. The authors claim that the framework can scale-in and out as required but there are not analytical results presented in the paper to support the claim. Proof-of-concept was shown to the IETF meetings but no results are presented in the paper.

A high-level architecture is also presented in \cite{Ziegler.iccws16}, where a network virtualization layer is laid over the physical infrastructure and under the service enablement layer which seems to be very close to control plane in concept. Architecture is using cognitive and cloud optimization domain and is named as CONE architecture. As it is also mentioned in the paper, many details are left out of the paper and only preliminary design is discussed. 

The proposed SDN-based mobile network architecture in \cite{Costa-Requena.eucnc15} moves EPC to cloud to benefit from NFV while maintaining 3GPP interfaces with legacy nodes. It suggests moving some functionalities of eNodeB, if required, to the cloud but the prototype uses the unchanged eNodeB. The proposed architecture is deployed over a testbed using two eNodeB from Nokia, OpenFlow enabled MPLS switch, open source S/P-GW and probing tools. The results show downtime of up to 2 seconds in live migration of a MME running on a VM instance.

Mobileflow \cite{Pentikousis2013} is one of the earliest solutions to resolve the compatibility issue between SDN-based networks and legacy systems using virtualization techniques. OpenRAN \cite{Yang.sigcomm13, Yang2015} is also a high level idea of a SDN-based mobile network through virtualization. A SDN controller would first create virtual RRU and virtual BBU using the physical resources and then dynamically optimize them according to the application. 

\cite{Gomez.noms14} develops an interesting concept of Hybrid-eNodeB (HeNB) which hosts a virtualized EPC (vEPC) along with all eNodeB functionalities. vEPC is the most fundamental version of EPC without the complexities of detailed control. vEPC needs the connections to physical EPC, although not all the time, to get the necessary information about billing, roaming, etc. HeNB greatly simplifies the deployment of 4G LTE and even allow standalone and self-powered UAVs (Unmanned Aerial Vehicles) and LAPs (Low Altitude Platforms) to host HeNB. \cite{Yousaf.icc14} presents the idea of virtualized EPC using a general purpose node (GPN). GPN is a core-class server with a hypervisor to provide virtual instances of EPC. The main purpose is to reduce cost through improvement in resource utilization. EPC nodes and interfaces remain unchanged in this work along with GTP tunneling.

{In a recent publication {\cite{Kitindi2017}}, the authors presented an architecture for WNV (Wireless Network Virtualization) which is similar to NFV concept but only for access networks. Their architecture places the virtualization layer or hypervisor as southbound interface between the physical network components or the data plane and the virtual resource management or the control plane. The virtual network functions are placed in the application plane. The paper also listed the outstanding issues and advantages of such an architecture.}

The proprietary virtualized access (vAccess) development platform from Freescale \cite{Rouwet2015} is created to facilitate the development of SDN/NFV products. It is built over OpenStack cloud computing platform with hardware accelerators for L1 real-time processing and Linux patch to support bounded guarantees for application start time. The vAccess platform provides the basic structure to build LTE base-station software.

In 2013, Huawei \cite{Huawei2013} unveiled SoftCom strategy. SoftCom is a holistic approach to network architecture based on cloud computing, SDN, and virtualization. The four key elements of SoftCom are cloudification of equipment through separation of hardware and software, cloudification of network through separation of forwarding and control, transforming traditional infrastructure to a cloud infrastructure, and transforming Telecom-oriented systems to Internet-oriented systems. For RAN, SoftCom is focusing on moving the control plane for small cells to macro base-station and centralization of BBU processing as in CloudRAN. SoftCom promises to develop an eco-system for simplification of network operations and innovation of open business models.

\subsection{{SDN Applications}}

{In this section, the discussion about application of SDN in improving and optimizing network management is included only to admire how the SDN research space is expanding and focus is turning towards details. There is huge amount of work in this respect specially since 2014, and it is outside the scope of the present paper to provide a comprehensive survey on applications.}

\subsubsection{{Resource Sharing}}

{A framework for co-primary spectrum sharing among MNOs (Mobile Network Operators) is developed in {\cite{Spapis.iisa14}}. The framework is based on an SDN architecture integrating Mobileflow {\cite{Pentikousis2013}} and SoftRAN {\cite{Gudipati.hotsdn13}}. Co-primary spectrum sharing enables optimal resource utilization among different operators. Mobileflow  {\cite{Pentikousis2013}} is the core network part of this scheme while SoftRAN {\cite{Gudipati.hotsdn13}} is evolved to enable access network sharing. The big base-station concept of SoftRAN is used in the paper for all base-stations of all operators. The logically centralized controller then assigns the resources based on the resource management rules. The scheme does not go into the details of SLA (Service Level Agreements) between MNOs and does not identify the business incentives and benefits. Security issues are also left out of the paper.}

{Radio resource sharing is also discussed in {\cite{Costanzo.icc14}}, where virtualization of LTE eNodeB is used to facilitate lease on-demand the physical infrastructure and resources of one operator to the flows of another operator. The framework is named OpeNB (Open eNodeB) and it is based on SDN and OpenFlow technologies. Besides OpenFlow controller, another controller called OpeNB controller provides signaling and management control for resource sharing. A main controller provides hand shaking between networks of different operators. System-level simulations show improved performance in terms of packet drop rate and throughput.}

{Another proposal for SDN-based smart gateways to connect S/P-GW of multiple operators is presented in {\cite{Thyagaturu2016}}. Here a SDN orchestrator is responsible to manage multiple smart-gateways in order to provide demand-based uplink capacity to the connected small cells. The smart-gateways allow resource sharing among multiple operators through SDN orchestrator.}

\subsubsection{{Mobility Management}}

{In {\cite{karimzadeh.closer14}}, the authors looked at the requirements of DMM (Distributed Mobility Management) and how SDN architecture can satisfy them. They also showed a simple example of how DMM can be implemented in a virtualized LTE environment. Their follow-up paper also provides simulation results of SDN-based DMM implementation {\cite{Valtulina.gcwkshps14}}. Their performance evaluation results show that seamless mobility management is possible when X2 path is used between eNodeBs.}

{The benefits of SDN in mobility management with global network view and flow-based control are highlighted in {\cite{Kuklinski.gcwkshps14}}. The authors discussed mobility management in SDN architectures with different centralization levels. They identified issues regarding session continuity and scalability of handover and showed how mobility management can be easily integrated with autonomic management mechanisms to optimize resource utilization.} 

\subsubsection{{Multi RAT scenarios}}

{A SDN enabled traffic offloading scheme is proposed in {\cite{Amani.icc14}}. A SDN controller runs software based RRM (Radio Resource Management) and PCRF (Policy Charging and Rules Function) modules to determine the offloading from LTE to WiFi depending on the network state and flow requirements. If both networks are unable to satisfy traffic requirements, the packets are dropped. Simulation results show 15\% less packet drop with the SDN-enabled offloading scheme. RRM is responsible for updating the network state information and PCRF is responsible for policy enforcement. The paper shows a very important use case where SDN can provide performance improvement in a straight forward manner.}

{A detailed sub-optimal online solution for end to end routing in mobile networks based on SDN and cloud computing is presented in {\cite{Cao2017}}. The objective is to maximize the amount of traffic accepted over time. The problem is formulated and an online algorithm is developed and tested via simulations for multi RAT (Radio Access Technology) scenario. }

{A wireless/wired backhaul solution based on SDN is presented in {\cite{Hurtado-Borras.icc15}} for small cells. The prototype performance is compared against 802.11s and significant improvement is observed when channel conditions are used to decide which packets to transmit over the backhaul. Similar idea is also presented in {\cite{Santos.wowmom16}} where some details of the on-going work on implementation are also discussed. An SDN controller can adaptively power on/off small cells based on the demand and traffic metrics.}

\subsection{Outstanding Issues}\label{issues_architecture}

Recent years have seen massive growth in SDN related research but there are still many open questions. The main focus of the present paper is on the major challenges. In this section, we limit ourselves to the open issues and problems in architectural domain only. Summarizing the literature in the current section, we identified the following important issues which should be looked at by the research community.   

\begin{enumerate}
	\item {\bf Policy compression} PCRF policies and rules need a consistent aggregation mechanism to be able to translate into routing rule for commercially available and OpenFlow-enabled routers and switches. 
	\item {\bf Access network layout} CloudRAN seems to be the choice for SDN-based access network but the fronthaul capacity and latency requirements create big hurdles in the production deployment. 3GPP's LTE Release 12 also defines Phantom cells to realize the densification of base-stations. Thorough investigation is needed to find the optimal alternative for different operating scenarios.
	\item {\bf Signaling/data split} Separation of control signaling and data transmission is proposed to achieve huge energy efficiency in BCG2 architecture and Phantom cell concept (cf.\ Section \ref{lte-background}). We found only three papers which proposed to integrate the idea into a SDN-based mobile network architecture  \cite{Liu2014}. The 85-90\% saving potential, as reported in  \cite{Earthd3.3}, is significant enough to encourage more exploration in this area. While designing the architecture for a new system, it may not be difficult to adopt the signaling/data split paradigm.
	\item {\bf Level of centralization} Moving all functionalities to cloud may not be a desirable choice for most of the operating scenarios. One reason is the limitations posed by GPP in processing real-time signals. Another is the inhibitive cost of fiber deployment if it is not already available. This area is being explored, mostly in terms of cost, but comprehensive investigation is still missing which can also look at the partial centralization scenarios and quantify the negatives, such as, the inability to carry out CoMP etc.
	\item {\bf Scalability} Inter-cluster communication between the controllers responsible for clusters of base-stations has not been explored at all.
	\item {\bf GPP implementation} EPC control nodes are suggested to be implemented as applications but carrier grade implementations are needed to ascertain the realizability of the idea.
	\item {\bf Edge computing} An opposite concept to centralization, edge computing also brings promise of performance enhancement and cost savings. It is vital to bring the two ideas together to see how they interplay and how practical it is to develop a plastic architecture that could be optimized for a given operational scenario.
	\item {\bf Integrated SDN and NFV} This area needs more investigation. Specially with practical limitations, we may not see fully centralized and fully software based system at least in the beginning. How these limitations affect the realization of NFV would be important and in fact may guide the SDN designs and development.  
\end{enumerate}

\section{Related Work}\label{rel_work}

Although, the development in the area of SDN-based mobile networks is trailing behind the work in wired and fixed networks but there are still a number of survey papers available in this domain. Now the question is, why do we need one more? In this section, we present a brief summary of the available surveys and while they all target to explore some aspect of the complex problem and moves forward the collective understanding of the research community, none has focused on the issues delaying the realization of SDN-based mobile systems. As per our knowledge, this is the first attempt to provide a holistic picture of the state of the art with emphasis on the weakest links. 

The survey in \cite{Yang.springer2015}, although is relatively recent but, presents a case for SDN by highlighting the issues with the current networks, such as, heterogeneous access networks, network ossification, increase in cost for operator amid dwindling revenues, etc.,  and how SDN can provide solutions. The paper, then, summarizes work in SDN and NFV for mobile and wireless networks and identifies open technical issues, such as, heterogeneous system's support, decomposition of control functions from protocol layers, scalability, customized and open API development, virtual machine migration, etc. The paper also advocates the joint design of SDN and NFV. SDN provides unprecedented visibility into the network allowing virtualization functions to have a clean abstraction to slice the resources.

The survey in \cite{Jagadeesan2014} discusses all SDN related research in wireless domain, including wireless LAN, Cellular, mesh networks, sensor networks, etc. Three publications- SoftCell \cite{Jin.conext13}, CellSDN \cite{Li.ewsdn12}, and SoftRAN \cite{Gudipati.hotsdn13}- related to cellular networks are mentioned in the survey, not surprising for such an early survey paper. The paper mostly lists the publications for each class of wireless networks and identifies the respective advantages and technical issues, such as, latency and channel variability for power allocation.

An almost similar list of publications in cellular domain appeared in \cite{Tomovic2014}, i.e., SoftCell \cite{Jin.conext13}, SoftRAN \cite{Gudipati.hotsdn13}, Mobileflow  \cite{Pentikousis2013}, and OpenRadio \cite{Bansal.hotsdn12}. This paper discusses the benefits, SDN brings to the access networks, such as, virtualization, interference management, and mobile traffic management. The challenges and issues in realization of SDN-based mobile networks are also briefly discussed, such as, actual cost saving margins, scalability of controller, security, isolation between network slices, estimation of channel load for proper resource allocation, and handover between different service providers. Similar survey of  SoftCell \cite{Jin.conext13}, SoftRAN \cite{Gudipati.hotsdn13}, Mobileflow  \cite{Pentikousis2013}, and CROWD project \cite{Ali-Ahmed.sdn4fns13} can be found in \cite{Li.2014}. The paper also discussed open research problems such as SDN-enabled cross-layer MIMO and heterogeneous radio technologies.

The survey in \cite{Chen2015} has a brief section on architectural designs of SDN-based mobile networks and they have classified them according to their closeness to CloudRAN or mobile edge computing. This paper is rather an overview of different issues including a brief history, high-level discussion about business case and challenges and technical problems, and standardization activities in this space.

In a recent and most comprehensive survey of SDN and virtualization research for LTE mobile networks \cite{Nguyen2016}, the authors have provided a general overview of SDN and virtualization technologies and their respective benefits. They have developed a taxonomy to survey the research space based on the elements of modern cellular systems, e.g., access network, core network, and backhaul. Within each class, the author further classified the material in terms of relevant topics, such as, resource virtualization, resource abstraction, mobility management, etc. They have also looked at the use cases in each class. It is the most comprehensive survey one could find in the radio access network research relevant to SDN. The thrust of the survey is complementary to the present paper. If the readers want better understanding about the material covered in Section \ref{sdran}, they are recommended to read \cite{Nguyen2016}. On the other hand, the open challenges briefly discussed at the end of \cite{Nguyen2016} and the relevant work under each challenge are discussed in detail in the present paper.    

The survey in \cite{Haque.2016} covers the holistic wireless space including cellular, WLAN, mesh, sensor, and home networks. In cellular networks, the major contribution of this paper is to cover only the significant papers in RAN and core network architectures with the focus on the inclusion of virtualization in the design and use of OpenFlow. The present paper provides much deeper discussion about the work in cellular SDN space and a much comprehensive coverage of the research area.

A much recent survey \cite{Nguyen2017} covers the proposals for SDN/NFV architectures for EPC, i.e., core network only. The authors found three different approaches to re-architect EPC, i.e., virtualizing EPC (vEPC) using NFV, decoupling control and user planes in vEPC with SDN technology, and fully SDN realized core network. They then classified the proposals according to the four attributes, namely, 1- architectural  approach (evolutionary or revolutionary), 2-technology  adoption (full or partial adoption of SDN or NFV or integrated NFV/SDN technology), 3-  functional  implementation (migration to VMs, decomposition of functions, merging of multiple functions into single unit), and 4- deployment  strategy (distributed or centralized). 

Another recent survey \cite{Namari2017} briefly surveys all technologies and applications associated with 5G. The survey also touches upon SDN and only superficially covers some research work under the theme. A more in-depth analysis of some of SDN-based mobile network architectures, i.e., \cite{Bansal.hotsdn12,Yap.sigcomm10,Pentikousis2013,Gudipati.hotsdn13,Cho2014,Jin.conext13,Akyildiz2015,Li.ewsdn12,Ali-Ahmed.sdn4fns13}, are presented in \cite{Tayyaba.ccode17} in terms of ideas presented in the proposals and their limitations. The survey in \cite{Temesgene2017} looks at the proposals for softwarization and cloudification of cellular networks in terms of optimization and provisions for energy harvesting for sustainable future. The gaps in the technologies are also identified. All of the above mentioned surveys, however, have a broader scope than just SDN-based mobile network architecture and they have only looked at some SDN papers appropriate for the major theme of their survey papers.

\section{Summary}\label{conc}

In this paper, we argue that transforming the current mobile network infrastructure to a SDN-enabled architecture may take more time than expected. With 5G around the corner (5G NR (New Radio) non-standalone standard approved in December 2017 and full standard is expected in 2018), it is likely that we may just see some point solutions based on SDN rather than a full scale roll-out of the technology. We identified six major roadblocks in realization of SDN-based mobile networks, i.e, fronthaul, latency of general purpose platforms, backward compatibility, disruptive deployment, SDN specific security vulnerabilities, and clear and compelling business case. We also looked at the state of the solutions under each issue. The major {lessons learned from} our study are summarized in the following subsections.

\subsection{Fronthaul}

\begin{enumerate}
\item Fronthaul's capacity and latency requirements need optical fiber links between the remote radio units (RRU) and the centralized processing unit. This is perhaps the biggest hurdle facing SDN deployments in places where fiber is not available. Countries with rich fiber coverage are in an advantageous position.
\item Microwave can be used in place of optical fiber but only for low rate CPRI (upto 2.5Gbits/s, future advances may support up to 10Gbps\footnote{\url{http://www.cablefree.net/wirelesstechnology/4glte/cpri-front-haul-technology/}})
\item CPRI, an internal base-station protocol to transport digitized waveform from cabinet to antenna heads, is the most widely used transport protocol for RRU-cloud connection but it has the worst data transport efficiency e.g., a 20MHz LTE channel can carry up to 150Mbps in downlink but requires CPRI rate of 2.5Gbps \cite{Murphy2015}. 
\item Other than capacity, delay and jitter are important issues for fronthaul. According to \cite{Chitimalla2017}, delay should remain under 100 microsecond and jitter should be under 65 nanoseconds for CPRI. 
\item Functional split is a promising solution to cater for fronthaul capacity constraints. It is highly probable that layer 1 and part of layer 2 would remain at the remote antenna sites, i.e., at RRUs and the rest of the layered protocol would move to the cloud. There is no clear winner in this case. You loose multiplexing gain when move away from full centralization option. Operating conditions and fronthaul constraints are the main drivers to choose the split option. Chinamobile's NGFI \cite{Chih-Lin2015} is also a promising option in practical deployments.
\item Use of Ethernet in fronthaul with or without CPRI is also very likely in future as it is a mature transport technology with much better efficiency than CPRI. 
\item {The most important development in this regard is IEEE 1914 working group (NGFI). The working group has two active projects: IEEE 1914.1 is studying NGFI of Chinamobile and IEEE 1914.3 is looking at encapsulation of digitized radio signal (I/Q samples) into Ethernet frames for fronthaul transmission.} \end{enumerate}  

\subsection{Latency of GPP}

GPPs are not designed for real-time signal processing and the delays could be of the order of tens of microseconds \cite{Tan2011}. There has been significant efforts in implementing the base-station components in GPP. So far, the data rates up to 43.8Mbps are supported on a 20MHz channel by an implementation of LTE uplink Rx PHY in GPP by Microsoft \cite{Tan2011}. It is still a long way to go as far as 5G goals of 1000x capacity and 100x data rate are concerned. 

Hybrid designs with some hardware accelerators for highly-computational blocks, e.g., turbo coding may provide a practical approach to meet the targets. GPP latency is shown to have decreased with processor speed and number of processor cores. 

\subsection{Backward Compatibility}

A practical realization of SDN should be able to work with the legacy systems specially 4G systems. The major property to hold is to keep the LTE interfaces and GTP tunneling in the new system as well. From innovation point of view, it is not a good idea as with time better alternatives are also surfacing e.g., MPLS labeling provides a faster alternative than GTP tunneling. 
Moreover, there has not been any work on UE handover between the two systems and regarding their co-existence in the same geographical area. {The recently approved 5G NR non-standalone standard defines additional interfaces and an evolved LTE eNodeB, i.e., eLTE eNB, to provide backward compatibility to 4G systems {\cite{3gpp.23.2017}}.}
\subsection{Disruptive Deployment}

An evolutionary approach for deployment is considered better and acceptable to the network owners and operators as any disruption would cause huge revenue loss. The ideal deployment trajectory will consist of point solutions leading to the full SDN-based systems. As seen before, the LTE interfaces and GTP tunneling are necessary for interworking with legacy systems and are kept in all proposals for evolutionary deployments studied in this paper. Most of the ideas, however, still remains at a  conceptual level and prototype implementation are necessary to ascertain their workability.


\subsection{SDN Specific Security Issues}

Security is perhaps the most researched area in SDN domain. Although, the development in wired networks are more advanced than in the wireless networks. SDN brings additional vulnerabilities to the network but at the same time provide enormous ease in controlling the network as well as visibility throughout the system. Exploiting these properties, better security systems can be introduced and implemented. The salient points from our study are given below.

\begin{enumerate}
	\item Controller is the most critical component of the SDN network and its compromise would give tremendous access to the attacker. DoS or DDoS attack on the controller would also be extremely devastating. Distributed control plane is an important paradigm for reliable and scalable network control which should be explored in detail and the inter-controller interfaces should be defined.
    \item There is still lack of standardization for secure northbound and southbound interfaces. A dedicated link can connect controller to its switches but appropriate level of encryption is needed to provide secure communication. The monetary implications of dedicated link and feasibility should be studied. 
    \item Programmability of the system may lead to policy inconsistencies and configuration errors. There are real-time tools for policy control resolution but they have scalability issues while dealing with large applications or large networks.
	\item Independent experimental evaluations and comprehensive assessment of solutions are required to check them for robustness, interoperability with third-party applications, integration of virtualization, and fast and dynamic reconfiguration. 
 	\item GPPs should also be hardened so the weaknesses in the processors cannot be exploited to gain control over the network or damage it. Moreover the current controller implementations, i.e., Floodlight, OpenDayLight, POX, and Beacon, also have resilience issues and common application bugs are enough to crash them. 
 	\item Real-time tools are needed to verify complete logical isolation between the virtual network sharing the same physical resources.
  	\item It is shown in \cite{Scott-Hayward2016}, that there are no solution till date for data leakage and modification in SDN networks. However, encryption is sought to be a valid solution for SDN networks as in traditional networks but the physical separation between controller and switches requires very important control information to be pass over the communication links. Their security should be comprehensively tested for the available encryption standards. 
    \item Most of the development is focused on OpenFlow and should be extended for broader SDN deployments. Moreover, OpenFlow cannot handle switch mobility yet.
    \item Security research related to SDN-based mobile networks is lagging behind. Wireless shared medium, mobility between different access technologies and different operators' network, constrained computational capacity, and backward compatibility to legacy networks are termed as important security issues relevant only to mobile networks. 
\end{enumerate}

\subsection{Clear and Compelling Business Case}

Virtualization, use of GPP instead of specialized hardware, and resource pooling in the cloud are the key ideas behind CAPEX and OPEX savings in SDN networks. Actual savings depend on the operating conditions, such as, availability or non-availability of optical fiber fronthaul, traffic load, equipment cost etc. In some business studies, 68\% lower CAPEX and 67\% lower OPEX are reported when LTE EPC is virtualized. Although, virtualization is not dependent on SDN and immediate gains are possible when traditional methods are used for virtualization. SDN, however, provides a sustainable business model for long-term future.

Some analyst also believe that the cost reduction due to GPP is based on the assumption that specialized  hardware is over-priced which is not always true. Considering the cost of transforming telco companies to suit the new SDN paradigm, the cost reduction may not be as great as predicted in some studies.

However, this has also been pointed out clearly that the real benefit of SDN is not in the cost savings but it is in providing the fail-fast capability to the networks. Carriers with such capability, where the deployment cycle for new technologies can be cut down to hours from days, will have a unique edge over others and eventually nobody could survive without it.

The disruptive model of transformation of communication network with full softwarization and centralization is very unlikely to be followed as none of the saving promises and monetary benefits can justify them. An incremental approach where SDN is introduced through point solutions in an already erected network seems to be more plausible. 

\subsection{Current State of the Art}


In our summary of architectural developments in SDN-based mobile networks, we list the repeated themes in the literature, such as, EPC nodes are mostly featured as application plane modules, SDN-based mobile networks may be close to CloudRAN in design with partial centralization where full centralization is costly, distributed control plane is scalable, OpenFlow is preferred choice of researchers at least for now, and an integrated SDN/NFV design is more desirable from business point of view. Outstanding issues are discussed in Section \ref{issues_architecture}.

There is still a gap between the state of the art and the point where SDN would be tech-ready for production deployment. SDN provides a novel and challenging playing field for research community but it is important to get the priorities right and emphasis should be on solving the problems which matter the most. 

\section*{Acknowledgment}
Partially funded by the EC H2020-ICT-2014-2 project 5G NORMA (\texttt{www.5gnorma.5g-ppp.eu})

\bibliographystyle{IEEEtran}

\bibliography{5G_SDN_RAN}

\end{document}